\documentclass[11pt]{article}
\usepackage{graphics} 
\usepackage{anysize} 
\usepackage{graphpap}
\usepackage{amssymb}
\marginsize{3cm}{3cm}{2cm}{2cm} 

\title{Flat (001) surfaces of II-VI semiconductors: A lattice gas model} 
\author{Martin Ahr\footnote{Corresponding author. Phone: +49 (0)931 888 4908 Fax: +49
(0)931 888 5141 E-mail ahr@physik.uni-wuerzburg.de}, 
\ Michael Biehl \\ 
Institut f\"{u}r Theoretische Physik und Astrophysik \\
Julius-Maximilians-Universit\"{a}t W\"{u}rzburg \\ 
Am Hubland, D-97074 W\"{u}rzburg, Germany}

\begin{document}
\newcommand{\cdtxt}{$c(2\times2)_{\mathrm{Cd}}$ }
\newcommand{\cdtxo}{$(2\times1)_{\mathrm{Cd}}$ }
\newcommand{\tetxo}{$(2\times1)_{\mathrm{Te}}$ }
\newcommand{\zntxt}{$(2\times2)_{\mathrm{Zn}}$ }
\newcommand{\setxo}{$(2\times1)_{\mathrm{Se}}$ }
\setlength{\unitlength}{0.01\textwidth}
\maketitle

\begin{abstract}
We present a two-dimensional lattice gas with anisotropic interactions
which model the known properties of the surface reconstructions of
CdTe and ZnSe. In contrast to an earlier publication \cite{baksv01}
the formation of anion
dimers is considered. This alters the behaviour of the model
considerably. We determine the phase diagram of this model by
means of transfer matrix calculations and Monte Carlo simulations. We
find qualitative agreement with the results of various experimental
investigations. \\ \\ 
{\bf Keywords:} Equilibrium thermodynamics and statistical mechanics,
Monte Carlo simulations, Surface relaxation and reconstruction,
Surface thermodynamics (including phase transitions), Cadmium
telluride, Zinc selenide, Low index single crystal surfaces.
\end{abstract}

\section{Introduction \label{intro}}

Within the last years, potential technological applications of
electronic devices based on II-VI semiconductors \cite{si00} have
inspired basic research concerning surfaces of these materials. In
this context, various studies have addressed the properties of surface
reconstructions. Experimental studies have investigated which
reconstructions are present \cite{tdbev94,ntss00} and how the
reconstruction of the surface is influenced by parameters like
temperature and particle flux in an MBE environment
\cite{wewr00,dbt96,vadt96}. The majority of this work has been devoted
to CdTe and ZnSe, where a fairly complete qualitative overview over
the phase diagram has been gained. An overview over the properties of
CdTe can be found in \cite{ct97}. On the other hand, there have been
theoretical investigations of the reconstructions of CdTe
\cite{gffh99} and ZnSe \cite{gn94,pc94} using density functional
theory. In these studies, knowledge about the chemical bonding of
surface atoms and ground state energies of various reconstructions has
been gained.
 
Being based on quantum mechanics, density functional theory is
believed to be exact apart from approximations made in the
calculation. The computational burden of this method is comparatively
high, which restricts its practical applicability to systems
consisting of only a few atoms. Due to the periodicity of crystal
surfaces this is not a severe restriction if one is interested in
ground state properties of the system which however, are strictly
relevant only at zero temperature. At higher temperature, the
properties of the surface will be influenced by thermodynamic effects.
These are particularly important if phase transitions between
different reconstructions occur. Their theoretical investigation
requires the study of systems with a large number of atoms which is
beyond the scope of first principles methods or molecular dynamics
simulations using realistic empirical potentials.

Consequently, simplifying models are needed which preserve essential
features of atomic interactions and can be investigated with moderate
numerical effort. In many cases, two-dimensional lattice gases have
been used successfully to model atoms adsorbed on a singular crystal
surface or the terminating layer of such a crystal
\cite{lbadebt00,baksv01,s80,ksb82}. In spite of the conceptual
simplicity of such models the interplay of attractive and repulsive
short range interactions can result in highly nontrivial critical
behaviour and complex phase diagrams. In this paper, we will follow
this approach to model the reconstructions of $(001)$ surfaces of CdTe
and ZnSe, our main focus being on CdTe.

The outline of this paper is as follows: In section \ref{overview}, we
will give a short review of the known facts about the reconstructions
of the $(001)$ surfaces of CdTe and ZnSe. In section \ref{lgasbasics},
we introduce a lattice gas model which considers the occupation of Cd
sites and the dimerization of Te atoms and discuss its phase
diagram. We conclude with a comparison of the features of our model
with experimental results in section \ref{experimentcompare}.

\section{Surface reconstructions of CdTe and ZnSe \label{overview}}
\begin{figure}
\begin{picture}(100, 30)(0, 0)
\put(0, 0){\resizebox{!}{0.3\textwidth}{\includegraphics{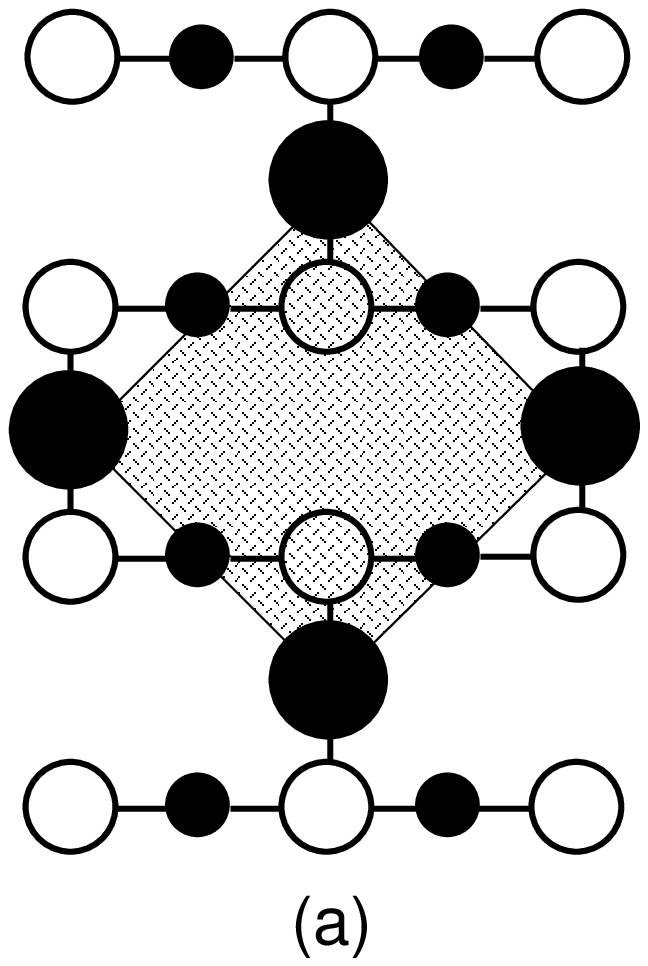}}}
\put(26, 0){\resizebox{!}{0.3\textwidth}{\includegraphics{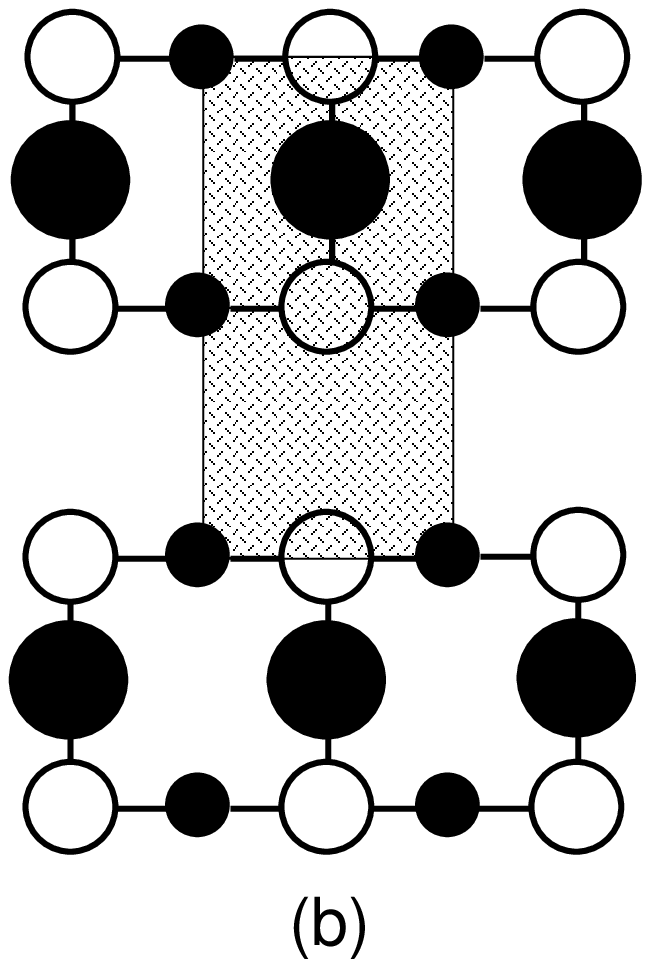}}}
\put(54,0){\resizebox{!}{0.3\textwidth}{\includegraphics{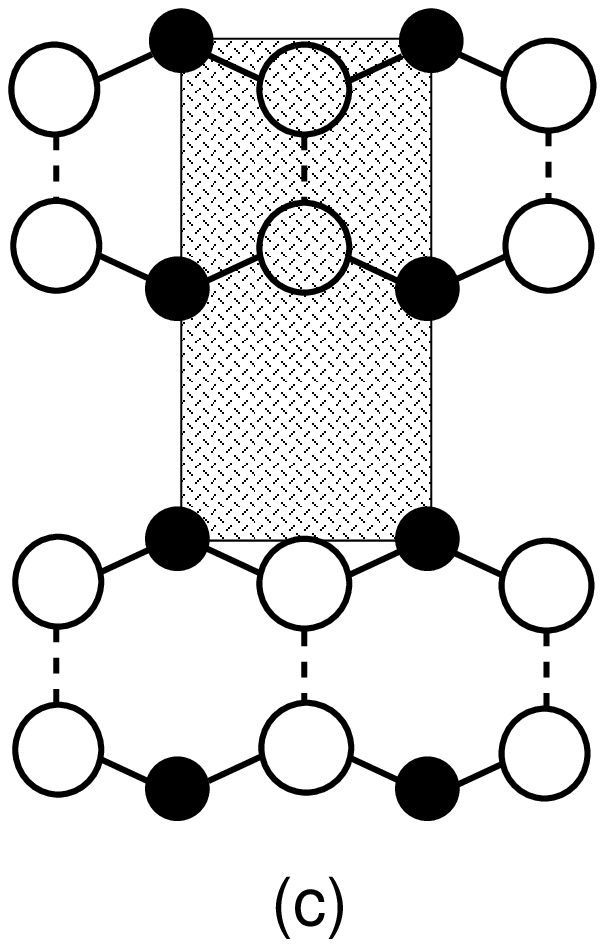}}}
\put(80,0){\resizebox{!}{0.3\textwidth}{\includegraphics{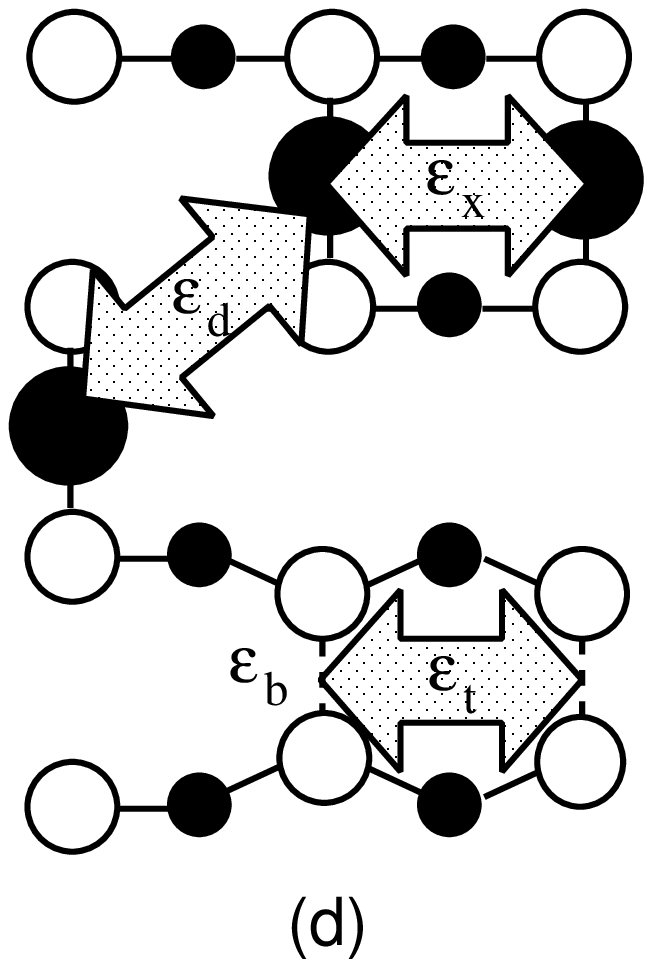}}}
\end{picture}
\caption{Panels (a),(b),(c): Sketches of the reconstructions of CdTe
which are discussed in this work. The grey rectangles show the surface
unit cells. The $[110]$ axis is aligned horizontally.  (a), (b):
Cd-terminated reconstructions. Panel (a) shows the \cdtxt
reconstruction, panel (b) the \cdtxo reconstruction. Panel (c) shows
the \tetxo reconstrution of surfaces terminated with a complete
monolayer of Te. Panel (d) shows the attractive couplings in our
model. \label{figrek}}
\end{figure} 
Both CdTe and ZnSe crystallize in the zinc-blende lattice. This
lattice structure is composed of alternating layers of cations and
anions which are parallel to the $(001)$ surface, such that an ideal
$(001)$ surface would be terminated by a complete layer of one
particle species. The positions of the atoms in one layer lie on a
regular square lattice with its axes oriented in the $[110]$
direction and the $[1\overline{1}0]$ direction.

Under vacuum, the $(001)$ surface of CdTe is Cd terminated. The
surface is characterized by vacancy structures where less than one
half of the potential Cd sites in the top layer are occupied
\cite{dbt96,vadt96}. This can be understood from simple
quantum-mechanical considerations like the electron counting rule
\cite{p89,h79}, which states that a surface terminated by a complete
layer of Cd is energetically unfavourable.  At low temperature, one
finds a \cdtxt reconstruction \cite{ct97,tdbev94}, where Cd atoms and
vacancies arrange in a checkerboard pattern (figure
\ref{figrek}a). Frequently, a contribution of a \cdtxo arrangement can
be found. In the \cdtxo structure, the Cd atoms arrange in rows along
the $[110]$ direction which alternate with rows of vacancies (figure
\ref{figrek}b). Density functional calculations \cite{gffh99} have
shown that the surface energies of the two competing reconstructions
are nearly degenerate and differ only by a small amount $\Delta E
\approx 0.016 eV$ per $(1 \times 1)$ surface unit cell. At a
temperature $T \approx 570 K$, there is a phase transition
\cite{tdbev94} above which the \cdtxo arrangement of the Cd atoms
dominates the surface. An analysis of high resolution low energy
electron diffraction (HRLEED) peaks \cite{ntss00} has shown, that
there is a high degree of disorder in this high temperature phase. One
finds elongated domains with a large correlation length of $\approx
375${\AA} in the $[1\overline{1}0]$ direction and a domain width as
small as $22${\AA} in the $[110]$ direction.

If CdTe is exposed to an external Cd flux in an MBE chamber, the
\cdtxt reconstruction is stabilized at temperatures above the
transition in vacuum.  Under a Te flux, the surface is Te terminated
with a \tetxo reconstruction. At small Te fluxes, the surface is
terminated by a complete monolayer of Te. The Te atoms on the surface
form dimers which arrange in rows (figure \ref{figrek}c). At high Te
flux and low temperature, one additional Te atom is incorporated into
each dimer, such that the Te coverage of the surface is 1.5. The
symmetry of this reconstruction is still $(2 \times 1)$, since the Te
trimers tend to arrange in rows.  A schematical phase diagram of the
surface can be found in \cite{ct97}.
 
Qualitatively, the properties of the reconstructions of ZnSe
\cite{ccd88,wewr00} are quite similar to those of CdTe, where the Zn
atoms are the counterparts of the Cd atoms and the Se atoms those of
Te, respectively. There is one important exception: to date, no $(2
\times 1)_{\mathrm{Zn}}$ reconstructed surface has been found at high
temperature. Density functional calculations \cite{gn94,pc94} yield a
higher energy difference $\Delta E \approx 0.03 eV$ per $(1 \times 1)$
surface unit cell between ideal $c(2\times 2)$ and $(2\times1)$
reconstructed surfaces, which is approximately twice the value
calculated for CdTe. As we will show, this greater energy difference
might explain the different behaviour of CdTe and ZnSe.

\section{The lattice gas model \label{lgasbasics}}

The basic structure of our model of (001) surfaces is the same for
different II-VI semiconductors.  The differences between the materials
are represented by the numerical values of parameters. In the
following, for simplicity we will loosely speak of Cd and Te atoms
instead of ``cations'' and ``anions'' without restricting ourselves to
a modelling of CdTe only.

In the lattice gas picture, a crystal lattice which is fixed in space
is considered. Each lattice site is either occupied by an atom or
empty. There are effective interactions between atoms which include
effects of surface strain and, at elevated temperature, lattice
vibrations. Therefore, there is no simple mapping between ground state
energies of the lattice gas model and surface energies determined from
density functional theory.

We model a flat $(001)$ surface of CdTe i.e. we neglect the influence
of steps on the reconstruction. This is a reasonable approximation if
the typical distance between steps is much greater than the size of
the unit cell of the reconstruction. In thermal equilibrium, this is
fulfilled at temperatures well below the roughening transition. In
this case, in the absence of bulk vacancies or other defects the
crystal is uniquely described by the state of the topmost monolayer
which consists of one Cd half-layer and a Te half-layer. In our model,
we consider such a monolayer with the Cd atoms on top. For simplicity,
we assume the Te layer to be fully occupied. This is not a severe
restriction, since the removal of Te atoms will uncover Cd atoms in
the layer below. Within the limit of a model of a flat surface, we
simply do not distinguish whether a Cd terminated surface has been
created by removing a Te layer or by depositing additional Cd atoms
onto an intact Te layer.

We use cartesian coordinates where the $x$-axis points in the $[110]$
direction, the $y$-axis points in the $[1\overline{1}0]$ direction and
the $z$-axis points in the $[001]$ direction. The origin of the
coordinate system is at the center of a Cd atom. Since we consider a
flat surface, the $z$-coordinate of all atoms of one species is
identical and will be omitted in the following.  Measuring the lattice
constant in appropriate units, the Cd atoms are at positions $(x, y)$
with integer $x, y$. The Te atoms are at positions $(x, y+ 1/2)$,
where $x, y \in \mathbb{Z}$. Tellurium dimers are created by the
formation of a chemical bond between neighbouring Te atoms in
$y$-direction ($[1\overline{1}0]$).  Since this is also the direction of
the bonds of a surface Cd atom, a Te dimer is formed by a pair of
atoms, which might also be the binding partners of a Cd atom above
them. This suggests a simple lattice gas representation both of the
occupation of Cd sites and Te dimerization: We consider a rectangular
array $x$ of integers $\{x_{i,j}\}_{i, j = 1}^{L, N}$. $x_{i,j} = 1$
represents a Cd atom at site $(i, j)$, which is bound to the Te atoms
at sites $(i, j-1/2)$ and $(i, j+1/2)$. $x_{i,j} = 2$ corresponds to a
dimerization of Te atoms $(i, j-1/2)$ and $(i, j+1/2)$. Otherwise,
$x_{i, j} = 0$.

In principle, one might also consider the formation of Te trimers by
introducing a fourth state $x_{i,j} = 3$. This corresponds to a trimer
which consists of the Te atoms at $(i, j-1/2)$ and $(i, j+1/2)$ and an
additional Te atom at $(i, j)$.  However, the formation of Te trimers
plays a role only if the surface is exposed to a strong flux of pure
Te at comparatively low temperature. Therefore, we have neglected this
effect to reduce the number of parameters of our model and the
numerical effort of its investigation.

\subsection{Interactions of atoms and dimers}

The detailed representation of the surface energies of a II-VI
compound certainly would require long-range interactions and terms
which depend on the simultaneous occupation of three or more
sites. However, it is plausible to assume that the dominant
contribution to the surface energy stems from pairwise interactions of
atoms at short distances. In this section, we introduce a Hamiltonian
which considers pairwise interactions between $x_{i,j}$ on nearest
neighbour sites and diagonal neighbour sites. These reflect the known
properties of the reconstructions of CdTe and ZnSe.

Due to the electron counting rule, surfaces with Cd coverages greater
than $1/2$ are unstable while both a \cdtxt reconstruction and a
\cdtxo reconstruction are permitted. This feature can be captured by
introducing a hardcore repulsion between Cd atoms on neighbouring
sites in the $y$-direction. In the $x$-direction, an attractive
interaction favours the occupation of nearest neighbour pairs the
strength of which is denoted by $\epsilon_x < 0$. An attractive
interaction $\epsilon_d < 0 $ between diagonal neighbours stabilizes
the \cdtxt reconstruction. These parameters are chosen such that the
energy difference $\Delta E = |2 \epsilon_d - \epsilon_x|/2$ between
these two reconstructions is small compared to the total surface
energy per site. 

The electron counting rule favours Te dimerization, but forbids the
formation of additional bonds of dimerized Te atoms. Therefore, we
forbid the simultaneous occupation of neighbour sites in the
y-direction with dimers (formation of chains of Te-Te-bonds) and Cd
atoms next to a dimer. These rules permit both a \tetxo
reconstruction, where the dimers arrange in rows and a
$c(2\times2)_{\mathrm{Te}}$ reconstruction, where they arrange in a
checkerboard pattern. Density functional calculations \cite{gffh99}
have shown that the surface energy of \tetxo is significantly lower
than that of a $c(2\times2)_{\mathrm{Te}}$ reconstruction. This may be
understood from the more efficient relaxation of surface strain in
\tetxo: Since the lattice is deformed in the same direction by dimers
on neighbouring sites in x-direction, such an arrangement will be
energetically favourable. We consider this fact by an attractive
nearest neigbour interaction $\epsilon_t$ between dimers in
x-direction. Additionally, a dimer contributes a binding energy
$\epsilon_b$ to the surface energy. Apart from the hardcore repulsion,
we consider no interaction between Cd atoms and dimers. An overview
over the couplings in our model can be found in figure \ref{figrek}d. 

This yields the Hamiltonian
\begin{equation}
H = \sum_{i,j = 1}^{L, N} \epsilon_x c_{i,j} c_{i+1,j} + \epsilon_d
c_{i,j} \left( c_{i+1,j-1} + c_{i+1,j+1} \right) + \epsilon_t d_{i,j}
d_{i+1,j} + \epsilon_b d_{i,j} - \mu c_{i,j},
\label{dreienergie}
\end{equation}
where we have introduced $c_{i,j} = \delta_{x_{i, j}, 1}$ and
$d_{i,j} = \delta_{x_{i,j}, 2}$. The boundary conditions are assumed
to be periodic. $c_{i, j}$ represents the occupation of lattice sites
with Cd atoms: $c_{i,j} = 1$ at the positions of Cd atoms and zero
otherwise. Similarely, the positions of dimers are given by $d_{i,j} =
1$. $\mu$ is the effective Cadmium chemical potential, which includes
the binding energy of surface Cd atoms to the substrate. The
groundstate of the system at $T = 0$ is determined by the chemical
potential $\mu$. For $\mu > \mu_0 = 2 \epsilon_d - \epsilon_t -
\epsilon_b$, the surface configuration with minimal energy is
\cdtxt. For $\mu < \mu_0$ the groundstate is \tetxo.

In the following, we measure energy in dimensionless units which have
been adjusted such that $\epsilon_d = -1$. Additionally, we set $k_{B}
= 1$ such that temperature is measured in units of $|\epsilon_d|$.

\subsection{Characterization of the surface configuration}

To characterize the surface reconstruction quantitatively, we evaluate
the mean Cd coverage $\rho_{Cd} = \left< c_{i, j} \right>_{i,j}$ and
the correlations
\begin{eqnarray}
C_{Cd}^{d} &=& \frac{1}{2} \left< c_{i, j} \left(c_{i+1,j+1} +
c_{i+1,j-1} \right) \right>_{i,j} \\ 
C_{Cd}^{x} &=& \left<c_{i,j} c_{i+1,j} \right>_{i,j}.
\end{eqnarray} 
$C_{Cd}^{d}$ measures the probability to find two diagonal neighbour
sites which are simultaneously occupied by Cd atoms. This is a measure
for the fraction of the surface which is covered by regions with a
{\em local} \cdtxt order.  Similarely, $C_{Cd}^{x}$ measures the
contribution of locally \cdtxo reconstructed regions in the system.

The {\em long range order} of the Cd atoms is measured by the order
parameters
\begin{eqnarray}
M_{Cd}^{(2\times 2)} &=& \frac{1}{L N} \sum_{i,j = 1}^{L,N} c_{i,j} \cos 
\left( \pi \left(i + j \right) \right) \\
M_{Cd}^{(2\times 1)} &=& \frac{1}{L N} \sum_{i,j = 1}^{L,N} c_{i,j} \cos 
\left( \pi j \right) .
\end{eqnarray} 
$M_{Cd}^{(2\times 2)}$ is the staggered magnetization of a system of
Ising variables $\{s_{i,j}\}_{i,j=1}^{L,N}$, where $s_{i,j} = 1$ if
$c_{i,j} = 1$ and $-1$ otherwise. Large absolute values indicate a
long range order of the \cdtxt reconstruction. Its counterpart
$M_{Cd}^{(2 \times 1)}$ measures the long range order of \cdtxo.
  
Further, we introduce the mean Tellurium dimerization $\rho_{D} :=
\left< d_{i,j} \right>_{i,j}$ and the dimer correlation $C_{D}^{x} :=
\left< d_{i,j} d_{i+1,j} \right>_{i,j}$, which characterizes the
number of dimers which are incorporated in {\em locally} $(2\times1)$
ordered areas. Their long range order is characterized by
\begin{equation}
M_{D}^{(2\times1)} = \frac{1}{L N} 
\sum_{i, j = 1}^{L, N} d_{i, j} \cos \left( \pi j \right) .
\label{mddef}
\end{equation}

\subsection{Methods of investigation}

We have investigated this model by means of the transfer matrix method
and Monte Carlo simulations. An introduction to the transfer matrix
technique can be found e.g. in \cite{cf72,na78}. This method allows
for a numerical calculation of the free energy of lattice systems with
short-range interactions. In general, the system size must be finite
in all directions but one. In our investigations, we choose the
$y$-direction as the infinite direction.  Since the computational
effort increases exponentially fast with the system size $L$ in the
$x$-direction, only comparatively small $L \leq 10$ are feasible.
However, this is sufficient to investigate the phase transitions of
the system in the limit $L \rightarrow \infty$ via finite size scaling
techniques.  To this end, we have applied the method suggested in
\cite{ber86} to determine the loci of first order phase transitions
and to compute coverage discontinuities.  Continuous transitions have
been investigated by means of phenomenological renormalization
group theory \cite{n82,na78}.

Additionally, we have performed Monte Carlo simulations of our
model. To obtain a reasonably fast equilibration, we have applied
continuous time algorithms \cite{nb99}. A more detailled description of
these methods is presented in the appendix.  We have simulated both
the grand-canonical ensemble where $\rho_{Cd}$ is controlled by $\mu$
and the canonical ensemble where the number of Cd atoms in the system
is fixed.  In general, the results of the transfer matrix calculations
and the Monte Carlo simulations are in good agreement (figure
\ref{fig1}).

\subsection{Results \label{results}}

To get insight into the typical behaviour of the model, we will first
present a detailed investigation using the parameter set $\epsilon_d =
\epsilon_b = -1$, $\epsilon_x = \epsilon_t = -1.9$. The choice
$\epsilon_x = -1.9$ reflects the fact that the energy difference
between \cdtxt and \cdtxo is small. The choice $\epsilon_b = -1$
yields low concentrations of Te atoms which are neither dimerized nor
bound to Cd atoms. This is consistent with experimental results.  For
simplicity, we have chosen $\epsilon_t = \epsilon_x$ as a typical
example of the case where both couplings are of the same order of
magnitude.  However, since there is no next-nearest neighbour
interaction between dimers, there is a significant difference in the
surface energies of \tetxo and $c(2\times2)_{\mathrm{Te} }$, which is
consistent with the results of \cite{gffh99}. 

In section \ref{parametersetinfluence} we discuss the influence of the
numerical values of the parameters on the phase diagram. We will show,
that a smaller value of $\epsilon_t$ affects the phase diagram
qualitatively. In particular, this concerns the high temperature
phases.

\subsubsection{Grand-canonical ensemble}
 
With the above parameter set, the Cd and the Te terminated
groundstates are separated by $\mu_0 = 0.9$.
\begin{figure}
\begin{center}
\begin{picture}(100, 66)(0, 0)
\put(0, 66){\resizebox{0.48\textwidth}{!}{\rotatebox{270}{\includegraphics{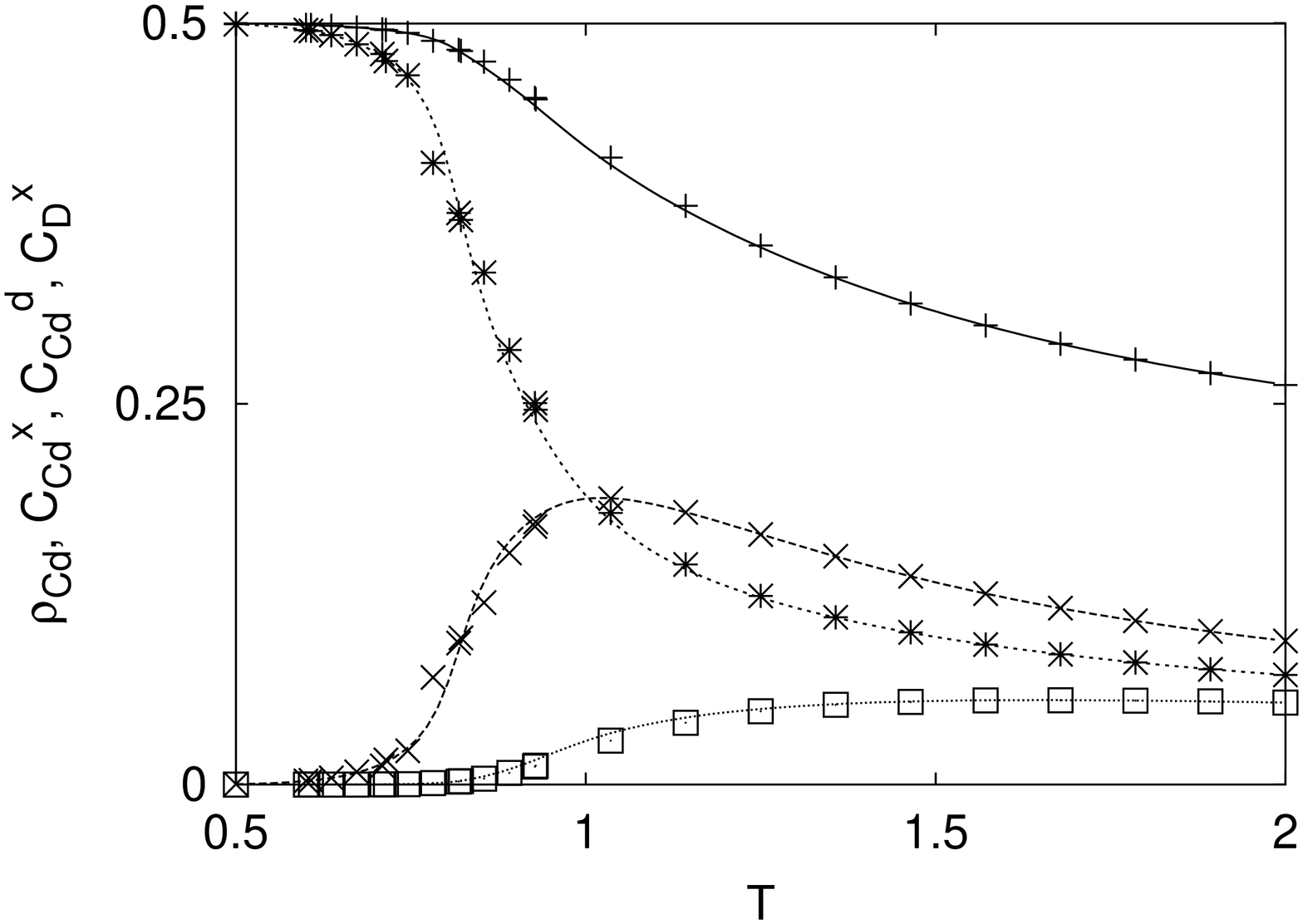}}}}
\put(50, 66){\resizebox{0.48\textwidth}{!}{\rotatebox{270}{\includegraphics{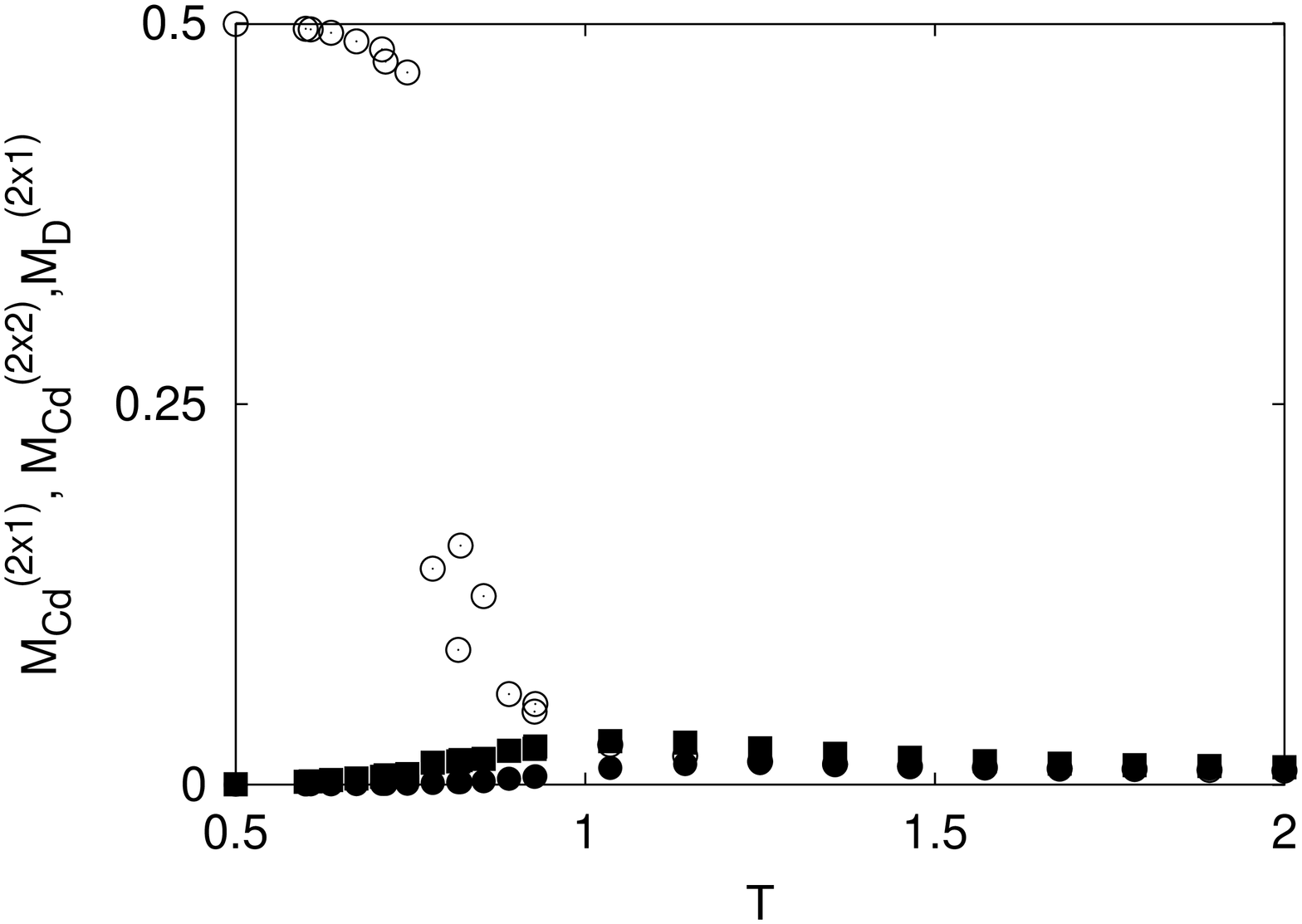}}}}
\put(0, 63){(a)}
\put(50, 63){(b)}
\put(0, 33){\resizebox{0.48\textwidth}{!}{\rotatebox{270}{\includegraphics{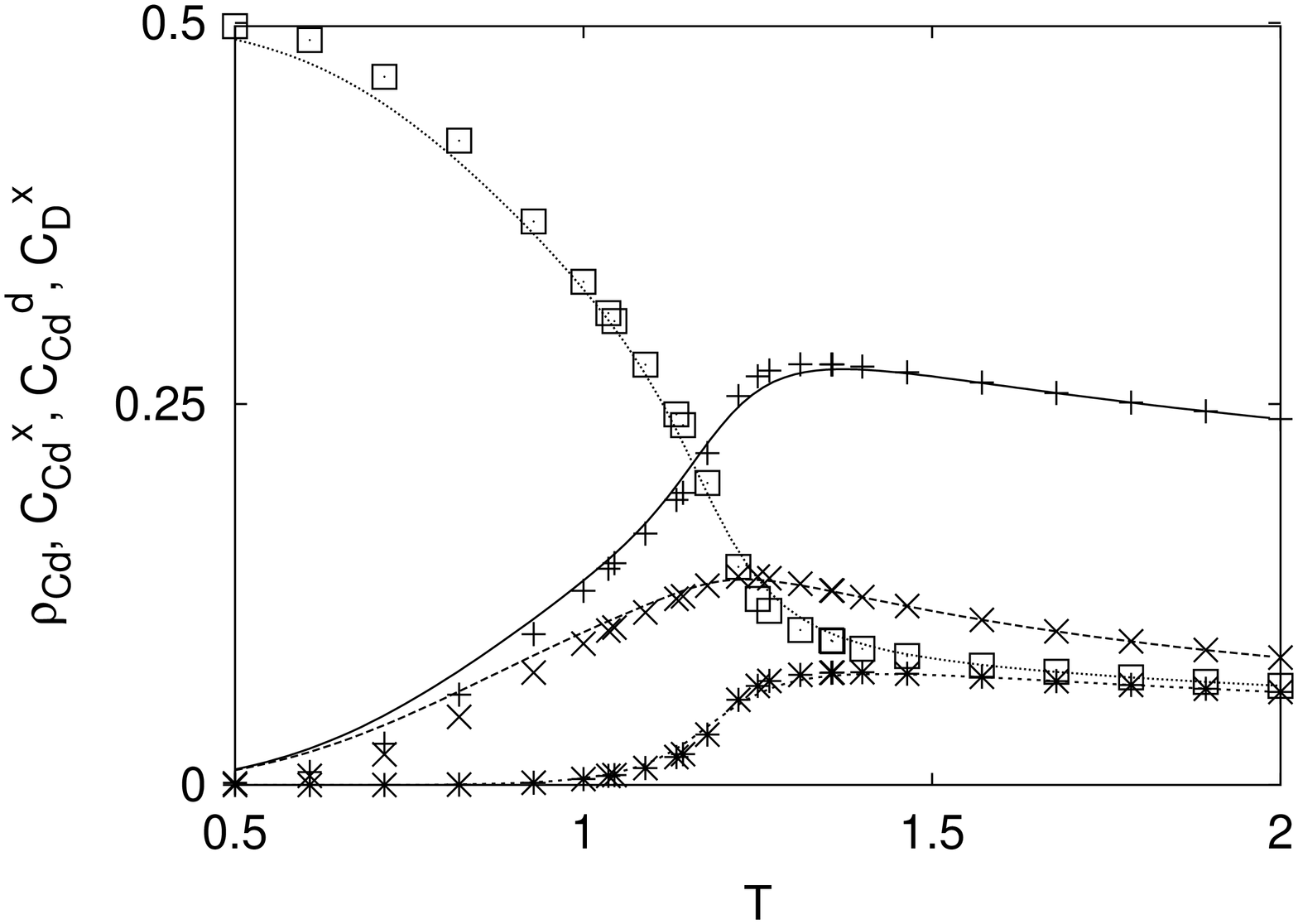}}}}
\put(50, 33){\resizebox{0.48\textwidth}{!}{\rotatebox{270}{\includegraphics{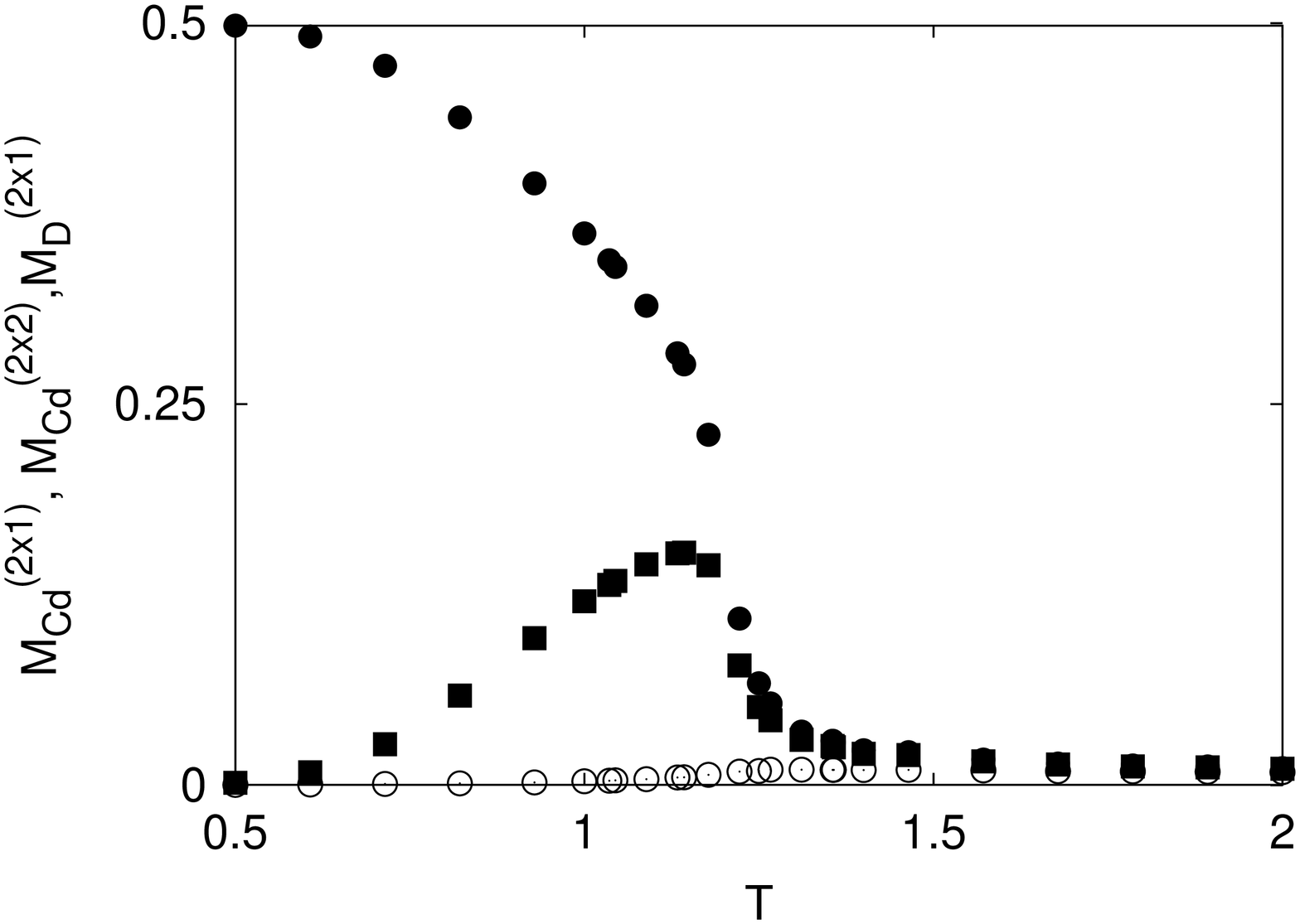}}}}
\put(0, 30){(c)}
\put(50, 30){(d)}
\end{picture}
\end{center}
\caption{Simulations at constant chemical potential. Panels (a), (b)
show the behaviour of the system with the couplings $\epsilon_d =
\epsilon_b = -1$, $\epsilon_x = \epsilon_t = -1.9$ and a cadmium
chemical potential $\mu = 1$. With these parameters, the ground state
of the system is \cdtxt. Panel (a) shows the Cd coverage $\rho_{Cd}\;
(+)$ and the correlations $C_{Cd}^{x} \; (\times)$, $C_{Cd}^{d}\;
(\ast)$ and $C_{d}^{x}\; (\boxdot)$. Panel (b) shows the mean absolute
of the order parameters $M_{Cd}^{(2\times2)}\; (\odot)$,
$M_{Cd}^{(2\times1)} \; (\blacksquare)$ and $M_{D}^{(2\times1)}\;
(\bullet)$.  The symbols show data from a simulation run at a system
size $L = N = 64$.  $4 \cdot 10^4 \cdot L N$ events have been
performed both for equilibration and for measurement. The lines have
been obtained by means of a transfer matrix calculation with a strip
width $L = 10$. The data shown in panels (c), (d) have been obtained
at $\mu = 0.8$, where the ground state is \tetxo. All other parameters
are identical to those used in (a), (b). \label{fig1}}
\end{figure}
In figure \ref{fig1} the temperature dependent behaviour of the system
at constant chemical potential is shown for $\mu = 1$ (figure
\ref{fig1}a,b) and $\mu = 0.8$ (figure \ref{fig1}c,d), which are
examples for both cases. At $\mu = 1$, the ordered \cdtxt phase at low
$T$ manifests itself in a high Cd coverage $\rho_{Cd}$ and values of
the correlation $C_{Cd}^{d}$ and the order parameter
$M_{Cd}^{(2\times2)}$ close to $0.5$. In figure \ref{fig2}d a surface
snapshot in this phase at $T = 0.71$ is shown. As expected, the Cd
atoms arrange preferentially in a checkerboard configuration.  At a
temperature $T = 0.83$, there is a first order phase transition to a
disordered phase. This is indicated by a decrease of the order
parameter $M_{Cd}^{(2\times2)}$ and the correlation
$C_{Cd}^{d}$. Simultaneously, $C_{Cd}^x$ starts to increase. At $T =
1$, both lines cross such that the high temperature behaviour of the
system is dominated by a local $(2\times1)$ ordering of the Cd
atoms. The simulation data shown in figure \ref{fig1}a,b have been
taken from two independent simulation runs. The system was initialized
with a perfect \cdtxt configuration at the lowest temperature
investigated. Then, as successively higher temperatures were imposed,
the surface configuration was kept as initial state of the next
simulation. This reflects in a small hysteresis effect in
$M_{Cd}^{(2\times2)}$ due to the first order nature of the phase
transition.

At $\mu = 0.8$, we obtain a completely different behaviour. Since the
groundstate is a \tetxo reconstruction, at small $T$ we measure low Cd
coverages and values of $C_{D}^{x}$ and $M_{D}^{(2\times1)}$ close to
$0.5$. The most frequent thermal excitations are Cd adatoms, the
density $\rho_{Cd}$ of which increases with $T$. Figure \ref{fig2}c
shows a surface snapshot at $T = 0.93$. The Cd atoms preferentially
arrange in rows such that $C_{Cd}^{x} \gg C_{Cd}^{d}$.  These rows
adapt to the structure of the \tetxo reconstruction. This yields
nonzero values of the order parameter $M_{Cd}^{(2\times1)}$, which
indicate a global ordering of Cd atoms in a $(2\times1)$ arrangement.
However, the interactions between the Cd atoms themselves are
insufficient to stabilize this global order. Instead, it is purely
induced by the environment of the Te dimers.

At $T = 1.67$, there is a first order phase transition above which the
system is in a disordered phase similar to that found at $\mu =
1$. Remarkably, we observe high $\rho_{Cd} \approx 0.2$ at
temperatures slightly below the phase transition. At the phase
transition, $\rho_{Cd}$ jumps to an even higher value.  Above the
transition it decreases slightly with $T$.

\subsubsection{Phase diagram}

\begin{figure}
\begin{picture}(100, 63)(0, 0)
\put(0,63){\resizebox{0.48\textwidth}{!}{\rotatebox{270}{\includegraphics{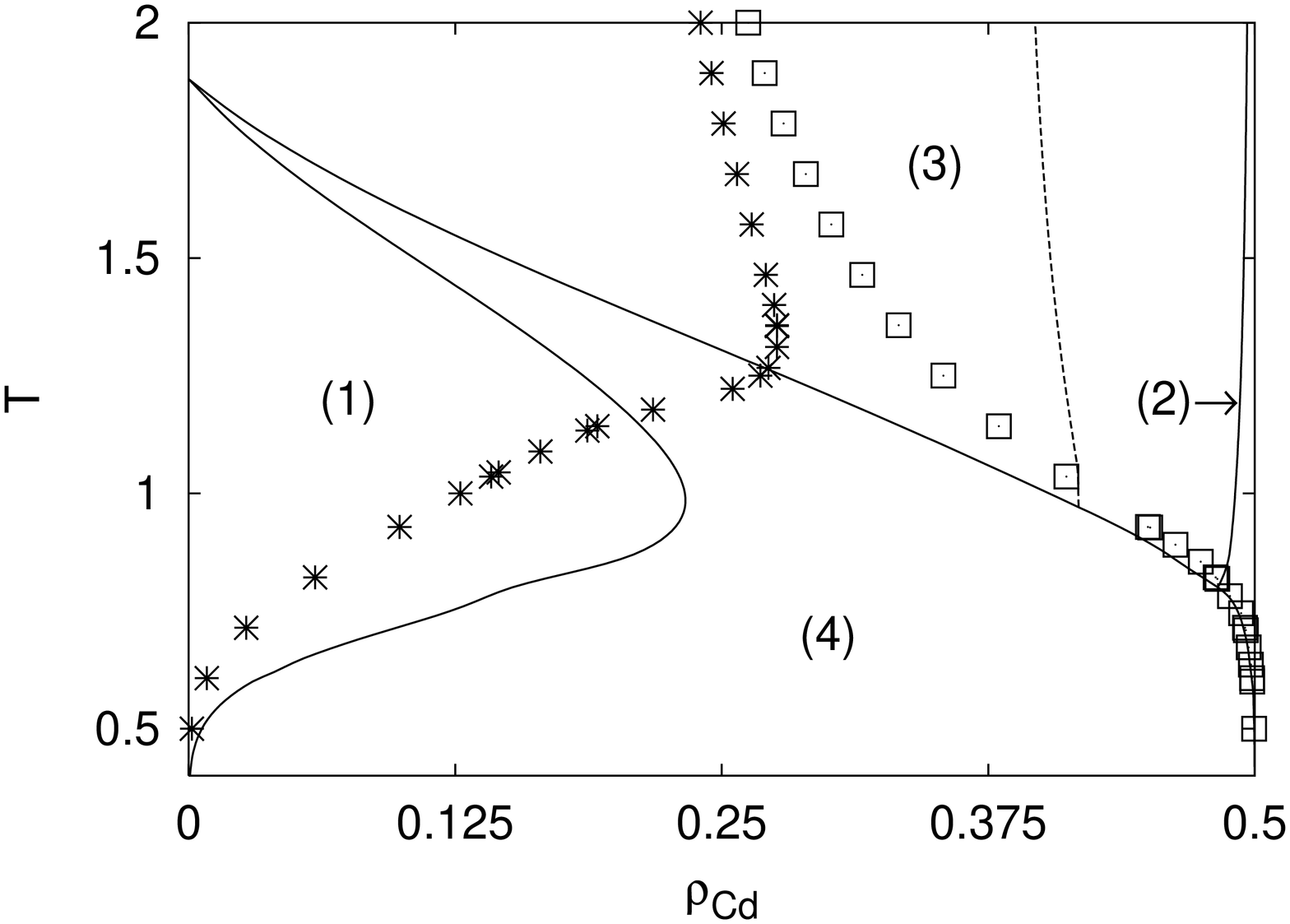}}}}
\put(50,63){\resizebox{0.48\textwidth}{!}{\rotatebox{270}{\includegraphics{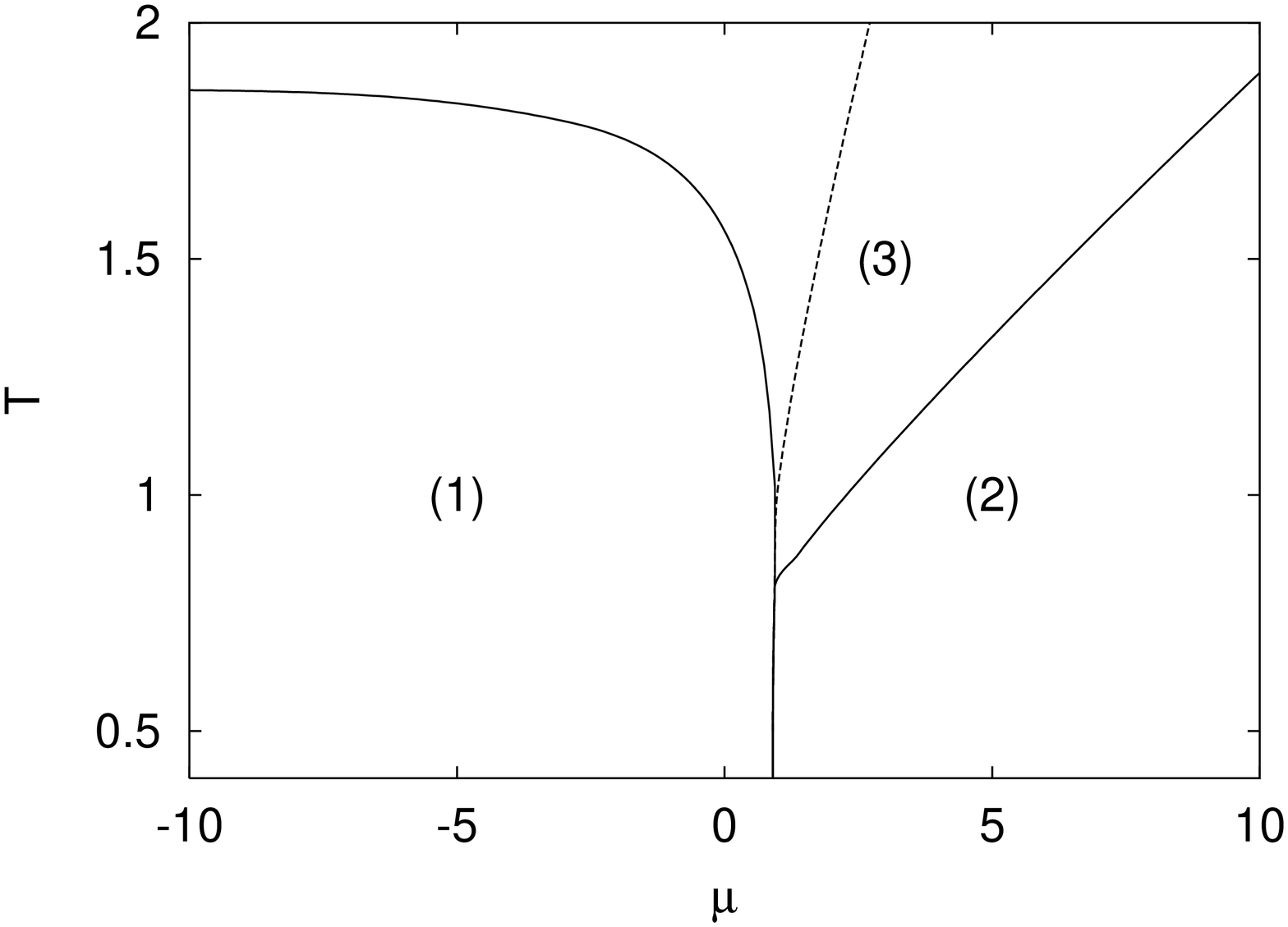}}}}
\put(0,60){(a)}
\put(50,60){(b)}
\put(0, 3){\resizebox{0.23\textwidth}{!}{\includegraphics{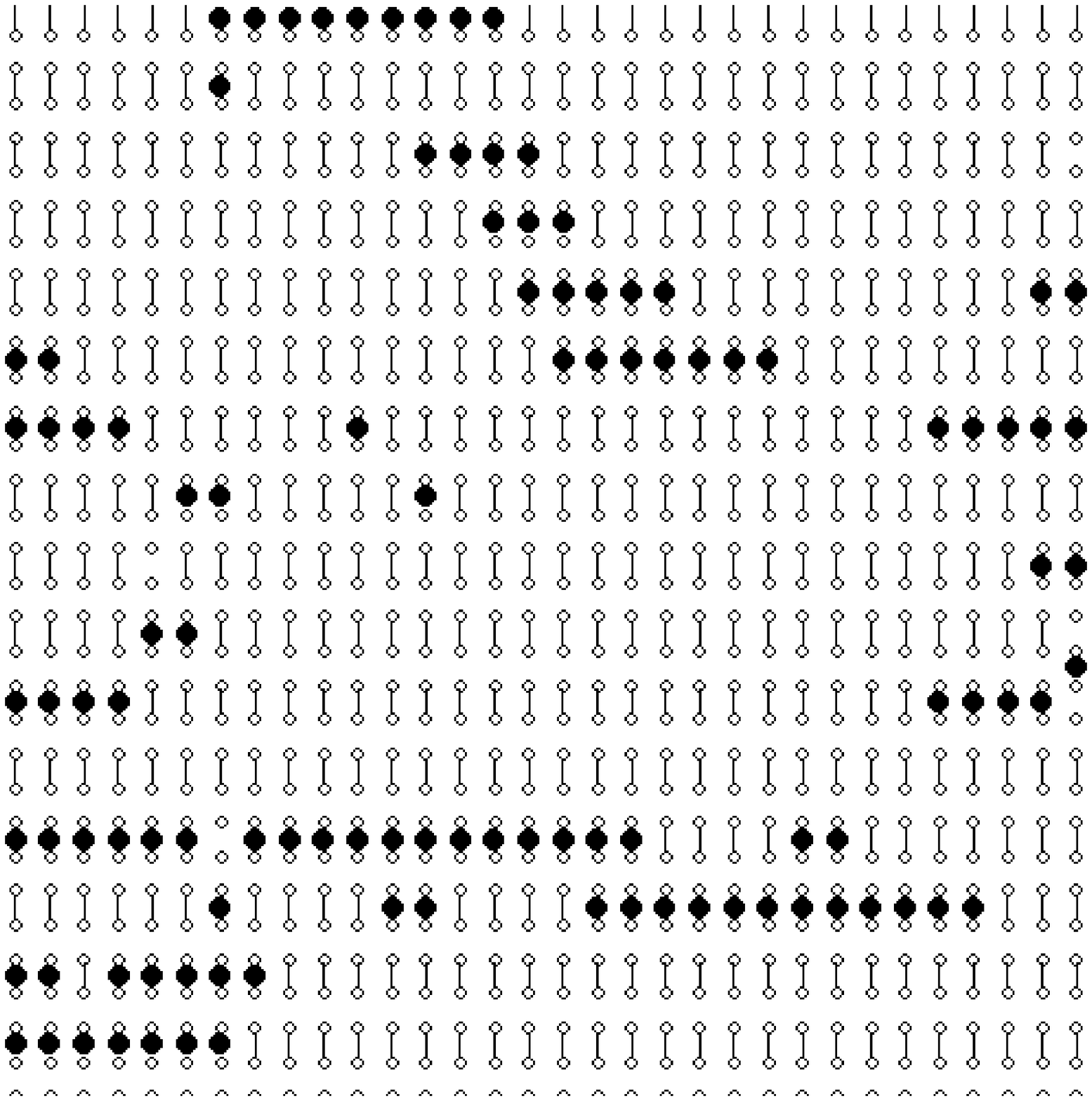}}}
\put(25, 3){\resizebox{0.23\textwidth}{!}{\includegraphics{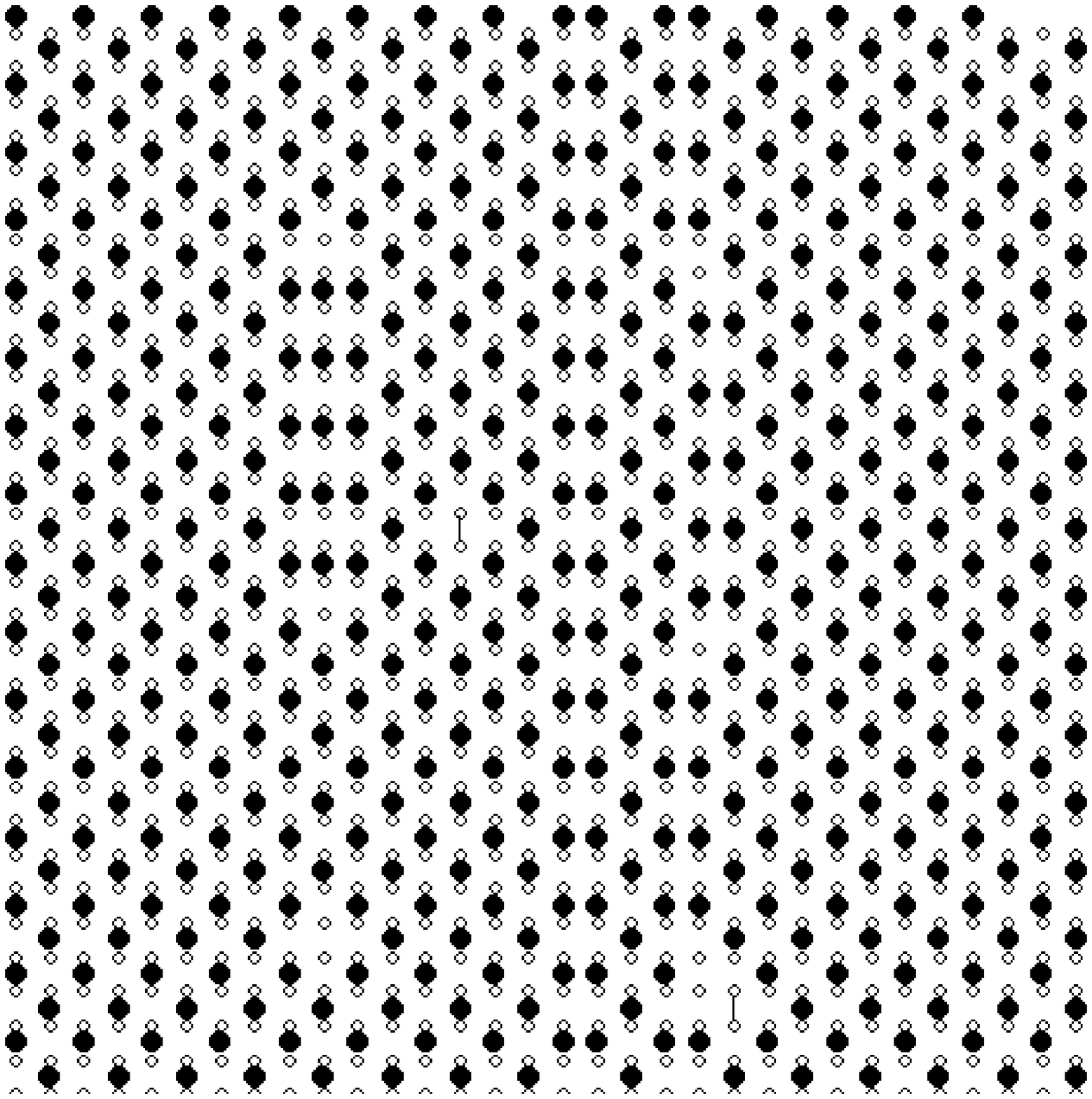}}}
\put(50, 3){\resizebox{0.23\textwidth}{!}{\includegraphics{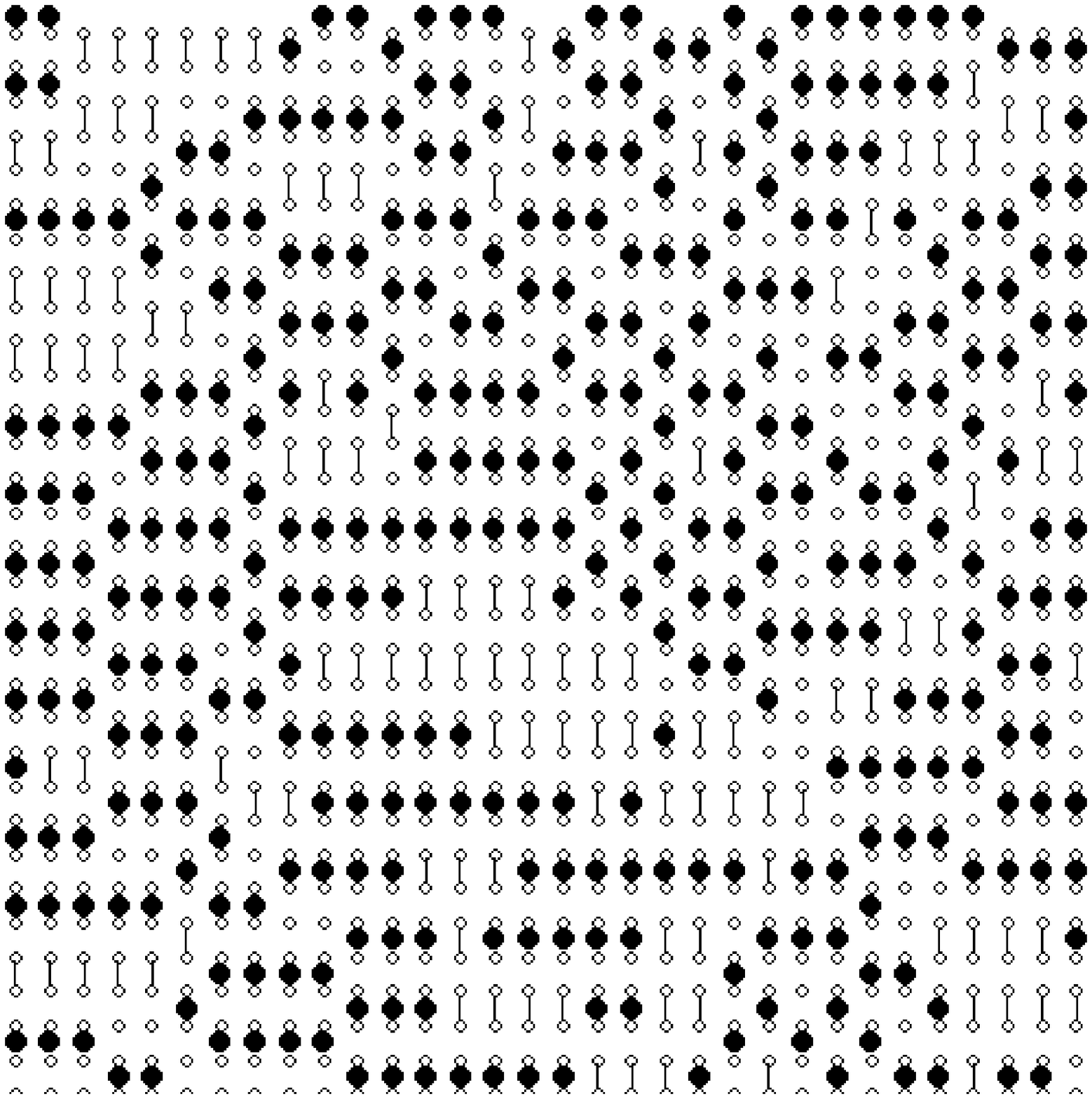}}}
\put(75, 3){\resizebox{0.23\textwidth}{!}{\includegraphics{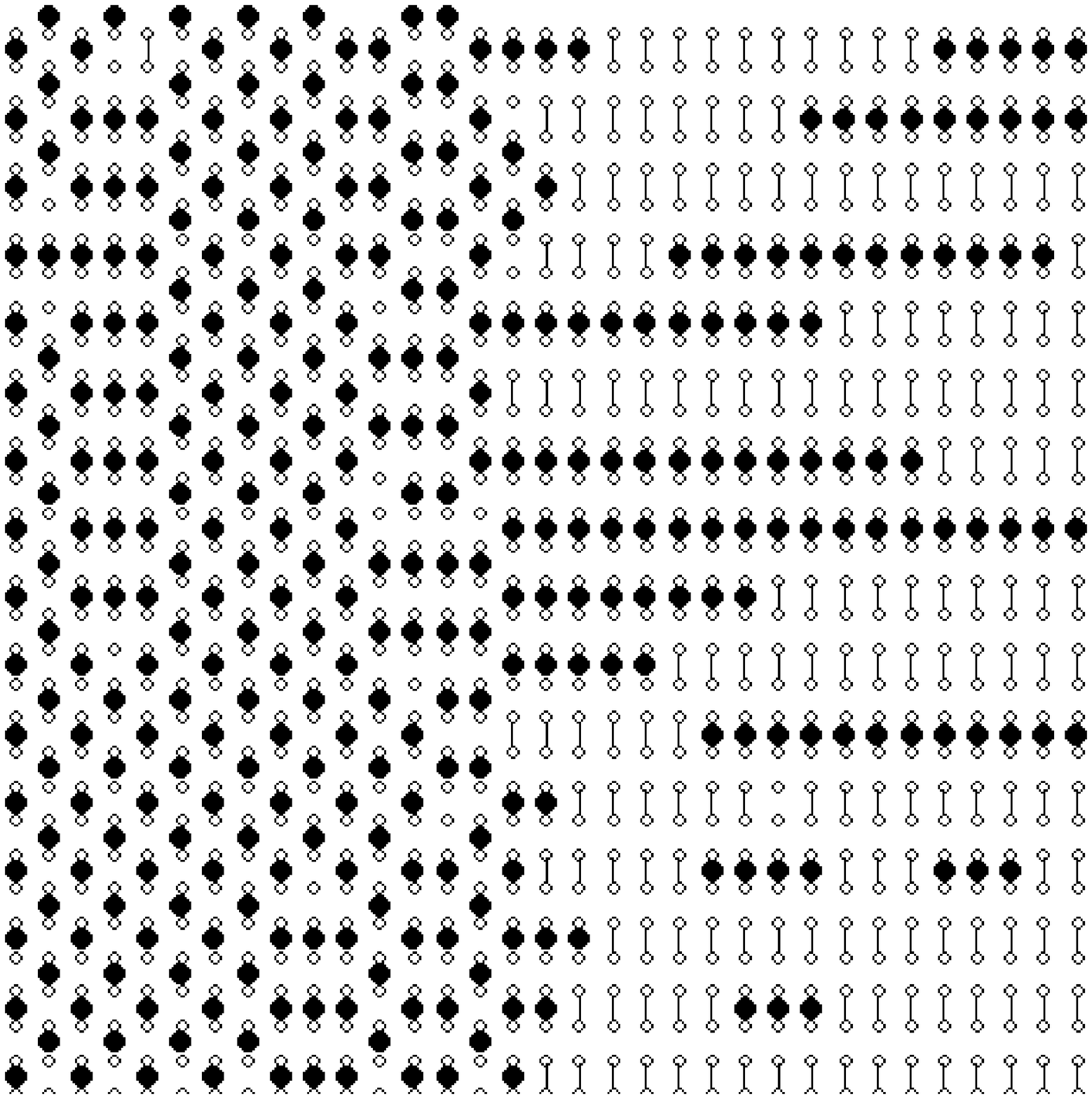}}}
\put(10, 0){(c)}
\put(35, 0){(d)}
\put(60, 0){(e)}
\put(85, 0){(f)}
\end{picture}
\caption{Phase diagram with parameters $\epsilon_d = \epsilon_b = -1$,
$\epsilon_x = \epsilon_t = -1.9$. The solid lines in panel (a) show
the lines of phase transitions as functions of $\rho_{Cd}$ and $T$, in
panel (b) they are shown in the $\mu$-$T$ plane. Note the offset in
the temperature axes. In region (1), the system is in a homogeneously
ordered phase with a \tetxo reconstruction, in region (2) it is
homogeneously ordered and \cdtxt reconstructed. Region (3) is the
disordered phase. On the left side of the dashed line, the Cd atoms
show preferentially a local $(2\times1)$ ordering ($C_{Cd}^{x} >
C_{Cd}^{d}$), while on its right side a $c(2\times2)$ ordering
dominates. Due to the coverage discontinuity between (1) and (2),(3),
there is a coexistence regime where regions with high and low
$\rho_{Cd}$ coexist (4). The symbols show lines of constant $\mu$
(Same data as in Figure \ref{fig1}a,c) Panels (c)-(f) show typical
surface snapshots. (c)-(e) correspond to grand-canonical
simulations. The snapshots show surface configurations after $8 \cdot
10^4 \cdot L N$ events. (c): $T = 0.93$, $\mu = 0.8$ (Phase 1), (d):
$T = 0.71$, $\mu = 1$ (Phase 2), (e)$T = 1.14$, $\mu = 1$ (Phase
3). (f) displays a surface configuration after $2 \cdot 10^4 \cdot L N
$ events in a canonical simulation at $T = 0.73$, $\rho_{Cd} = 0.35$
(Region 4, Coexistence of Phases 1 and 2).  All snapshots show sections
of $32\times32$ lattice constants of systems of size $L = N =
64$. \label{fig2}}
\end{figure}
Figure \ref{fig2}a,b shows the phase diagram of the model, which has
been extrapolated from transfer matrix calculations with strip widths
$L$ of 6, 8 and 10 lattice constants. In figure \ref{fig2}b, the lines
of phase transitions in the $\mu$-$T$ plane have been plotted. At low
temperature, the system is either in an ordered \tetxo (1) or an
ordered \cdtxt phase (2). The line of phase transition between these
phases starts at at zero temperature and $\mu = \mu_0$, where the
energies of both reconstructions are degenerate. For $0 < T < T_t =
0.84$, it remains at the same chemical potential $\mu_0$, apart from
small numerical uncertainties of extrapolation. In consequence, a
phase transition between a \cdtxt and a \tetxo reconstruction cannot
be observed at constant chemical potential. The disordered
phase (3) exists for $T \geq T_t$.  At the point $(T = T_t, \mu =
\mu_0)$ five phases coexist: the disordered phase, and \cdtxt and
\tetxo reconstructed phases in two sublattices, corresponding to
positive and negative values of the order parameters
$M_{Cd}^{(2\times2)}$ and $M_{D}^{(2\times1)}$, respectively. The
\tetxo reconstructed phase (1) exists only at temperatures below a
critical temperature $T_{c}^{1} = \epsilon_t$. At $T_{c}^{1}$, the
line of the phase transition to the disordered phase (3) diverges to
$\mu = - \infty$. On the contrary, the \cdtxt reconstructed phase (2)
may exist at arbitrary temperature, if the Cd chemical potential is
large enough.  The dashed line in figure \ref{fig2}b shows the
chemical potential, at which the correlations $C_{Cd}^{d}$ and
$C_{Cd}^{x}$ in the disordered phase are equal. For smaller $\mu$, the
local ordering of Cd atoms is dominated by a $(2\times1)$ arrangement,
while for larger $\mu$ they prefer a local $c(2\times2)$ ordering.

Figure \ref{fig2}b shows the phase diagram in the $\rho_{Cd}$-$T$
plane. At the transition between the phases (2) and (3), $\rho_{Cd}$
varies continuously. The phase transition from phase (2) to phase (1)
occurs at a temperature independent chemical potential
$\mu_0$. Therefore, there is no coverage discontinuity if temperature
is varied at a constant chemical potential $\mu > \mu_0$ where the
groundstate is a \cdtxt reconstruction.

On the contrary, there is a coverage discontinuity at the phase
transition between the ordered phases (1) and (2) and at the
transition between phases (1) and (3). These discontinuities yield a
coexistence regime (4). Here, for $T < T_t$, \cdtxt and \tetxo
reconstructed phases coexist, while for $T > T_t$ the \tetxo
reconstructed phase coexists with the disordered phase. Figure
\ref{fig2}f shows a surface snapshot from a simulation which was
performed at a constant Cd density $\rho_{Cd} = 0.35$ and a
temperature $T = 0.73$. This is a typical surface configuration in the
regime where the ordered phases (1) and (2) coexist. The system is
separated in two phases, one with a high $\rho_{Cd}$ and a \cdtxt
reconstruction, and another one with a \tetxo reconstruction and a low
concentration of Cd adatoms. The local values of $\rho_{Cd}$ in both
regions are given by the left and the right boundary of the
coexistence regime. At low temperature one obtains $\rho_{Cd} \sim
0.5$ in the \cdtxt phase and $\rho_{Cd} \sim 0$ in the \tetxo
phase. At temperatures $T \gtrapprox 0.6$, $\rho_{Cd}$ in the \tetxo
phase increases strongly with $T$ and obtains its maximal value
$\rho_{Cd}^{1,max} = 0.23$ at $T = 0.98$. At even higher temperature
it decreases with $T$ and becomes zero at $T_{c}^{1}$. On the
contrary, $\rho_{Cd}$ in the \cdtxt phase remains high for $T <
T_t$. At $T > T_t$, the Cd-rich phase is disordered. Then, $\rho_{Cd}$
at the right boundary of the coexistence regime decreases with $T$.
At $T_{c}^{1}$, it becomes zero and the coexistence regime
disappears. The dashed line in figure \ref{fig2}a marks the values of
$\rho_{Cd}$ at which $C_{Cd}^{x} = C_{Cd}^{d}$. For smaller coverages,
$C_{Cd}^{x} > C_{Cd}^{d}$ such that the local ordering of the Cd atoms
is dominated by a $(2\times1)$ arrangement, while for greater
coverages, $C_{Cd}^{x} < C_{Cd}^{d}$. The values of $\rho_{Cd}$ which
have been measured in the simulations shown in figure \ref{fig1} are
plotted in the phase diagram as examples of lines of constant chemical
potential. The loci of the phase transitions and the point where
$C_{Cd}^{x} = C_{Cd}^{d}$ (at $\mu = 1$) are in good agreement with
the results of the transfer matrix extrapolation.
 
\subsubsection{Canonical ensemble \label{canonicalresults}}

\begin{figure}
\begin{picture}(100, 33)(0, 0)
\put(0,33){\resizebox{0.48\textwidth}{!}{\rotatebox{270}{\includegraphics{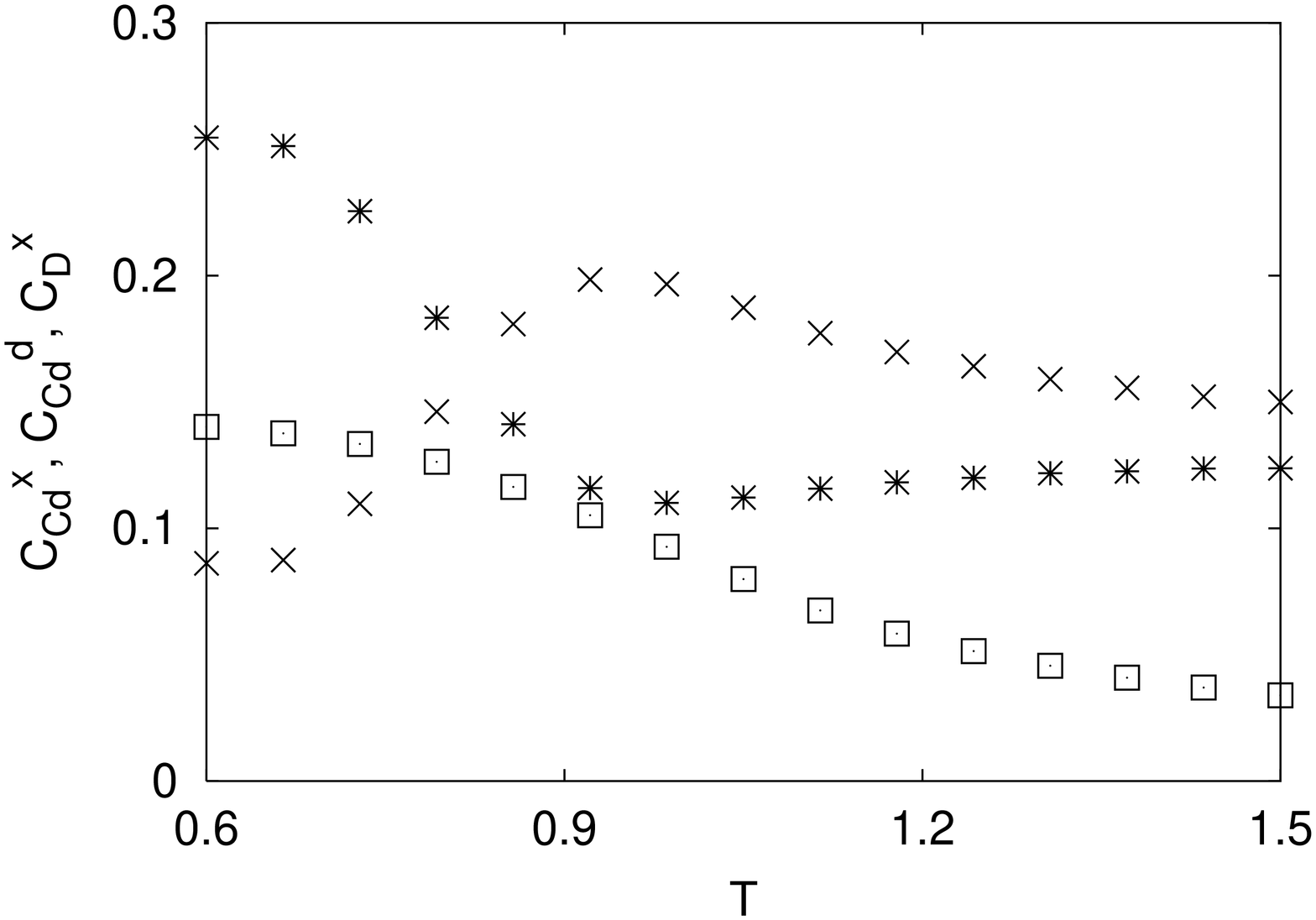}}}}
\put(50,33){\resizebox{0.48\textwidth}{!}{\rotatebox{270}{\includegraphics{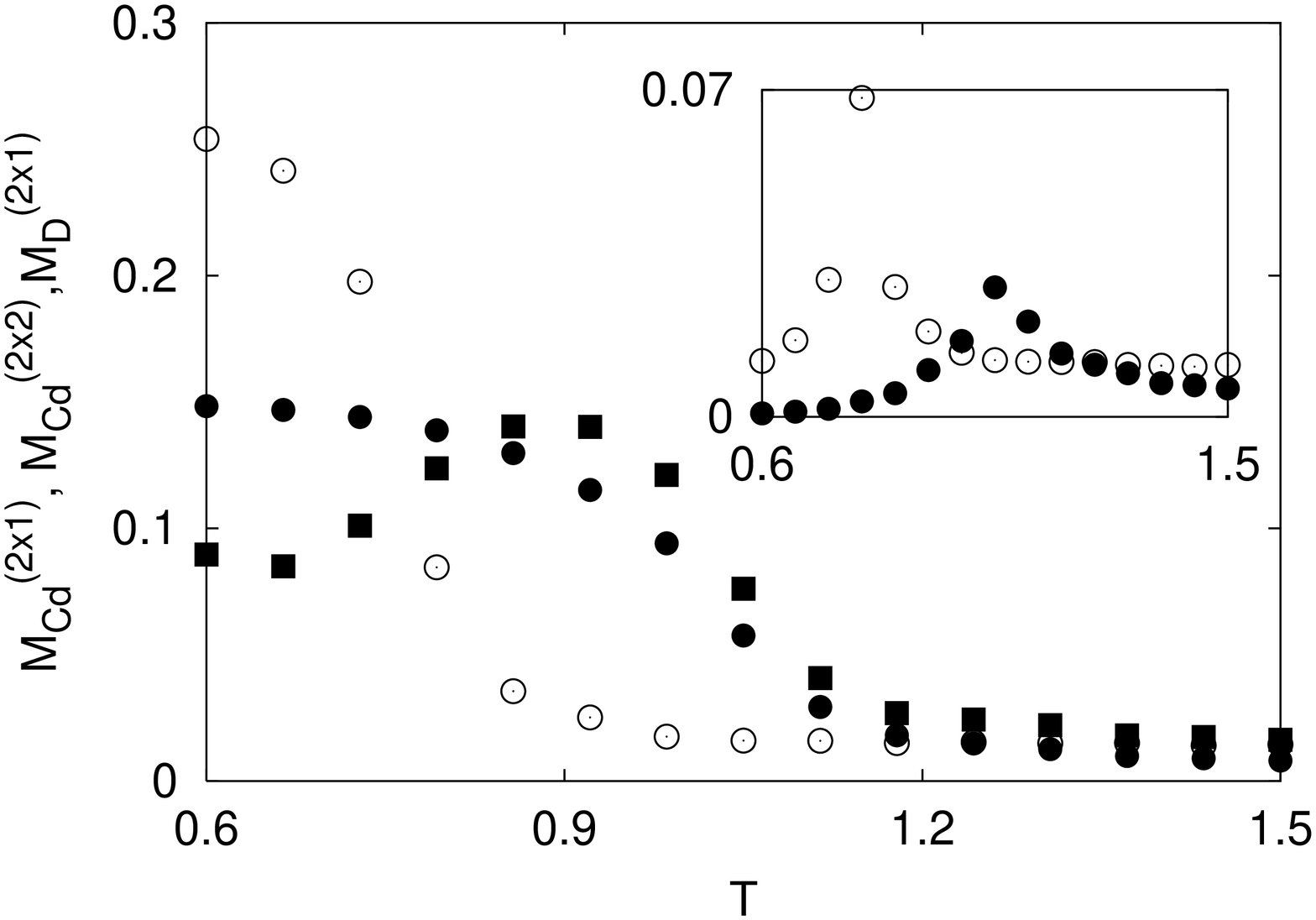}}}}
\put(0,30){(a)}
\put(50,30){(b)}
\end{picture}
\caption{Results of a simulation of a $64\times64$ system at conserved
$\rho_{Cd} = 0.35$ and couplings $\epsilon_d = \epsilon_b = -1$,
$\epsilon_x = \epsilon_t = -1.9$. $10^4 \cdot L N $ events have been
performed both for equilibration and for data sampling. Panel (a)
shows the correlations $C_{Cd}^{x} \; (\times)$, $C_{Cd}^{d}\; (\ast)$
and $C_{d}^{x}\; (\boxdot)$.  Panel (b) shows the mean absolute of the
order parameters $M_{Cd}^{(2\times2)}\; (\odot)$, $M_{Cd}^{(2\times1)}
\; (\blacksquare)$ and $M_{D}^{(2\times1)}\; (\bullet)$. The inset
shows standard deviations of order parameters. The meaning of the
symbols is the same as in the large picture. \label{fig3}}
\end{figure}

The temperature dependent behaviour of the model at a constant Cd
adatom density $\rho_{Cd} = 0.35$ is shown in figure \ref{fig3}. This
is an example which shows the typical behaviour of the model under the
conditions of phase separation. At low temperature, most of the Cd
atoms are concentrated in a Cd rich phase with a \cdtxt
reconstruction. The remaining area of the system is covered with a
\tetxo reconstructed phase. Since both phases are long-range ordered,
we obtain large values of the order parameters $M_{Cd}^{(2\times2)}$
and $M_{D}^{(2\times1)}$ and the corresponding correlations. The
fraction of Cd atoms which occupy sites in the Te rich phase yields
nonzero values of the order parameter $M_{Cd}^{(2\times1)}$. As
temperature increases, more and more Cd atoms pass into the Te rich
phase. This yields an increase in $C_{Cd}^{x}$ and
$M_{Cd}^{(2\times1)}$ and a decrease of $C_{Cd}^{d}$ and
$M_{Cd}^{(2\times2)}$ due to the $(2\times1)$ ordering of the Cd atoms
{\em in the Te rich phase}. At the temperature $T_t$, the Cd rich
phase undergoes the order-disorder transition where
$M_{Cd}^{(2\times2)}$ drops to zero. At this phase transition, the
fraction of Cd atoms which are incorporated in locally $(2\times1)$
ordered configurations {\em in the Cd rich phase} increases. Thus,
there are two independent effects which lead to a dominance of
$C_{Cd}^{x}$ over $C_{Cd}^{d}$ at higher temperature: The specific
shape of the left boundary of the coexistence regime and the
order-disorder transition of the Cd rich phase. At a temperature
$T_c^4$, the system is leaving the coexistence region into phase
(3). At this phase transition, the separation between a Te rich and a
Cd rich phase disappears, such that at high temperature the system is
in a homogeneous, disordered state.  The order parameters
$M_{D}^{(2\times1)}$ and $M_{Cd}^{(2\times1)}$ become zero and the
local correlation between dimers, $C_{D}^{(2\times1)}$, decreases. In
the simulations we have determined the critical temperatures of the
phase transitions from the standard deviations of the order parameters
$M_{Cd}^{(2\times2)}$ and $M_{D}^{(2\times1)}$. $M_{Cd}^{(2\times2)}$
is peaked at $T_t$, where the long range order of the Cd atoms is
lost, while $M_{D}^{(2\times1)}$ is peaked at $T_{c}^4$, where the
system leaves the coexistence regime and the dimers lose their long
range order. We obtain $T_t = 0.79 \pm 0.2$ and $T_c^4 = 1.0 \pm 0.2$,
where the uncertainty is due to the temperature spacing between the
single simulations. These values are systematically lower than the
theoretical results of the transfer matrix extrapolation ($T_t =
0.84$, $T_c^4 = 1.11$). However, the theoretical results are valid in
the limit of an infinite system size, where the free energy of the
phase boundary can be neglected compared to that of the bulk of the
phases. We have verified that the systematic deviation between theory
and simulations decreases with the system size.

\subsubsection{The influence of the parameter set on the phase diagram
\label{parametersetinfluence}}

\begin{figure}
\begin{picture}(100, 33)(0, 0)
\put(0,33){\resizebox{0.48\textwidth}{!}{\rotatebox{270}{\includegraphics{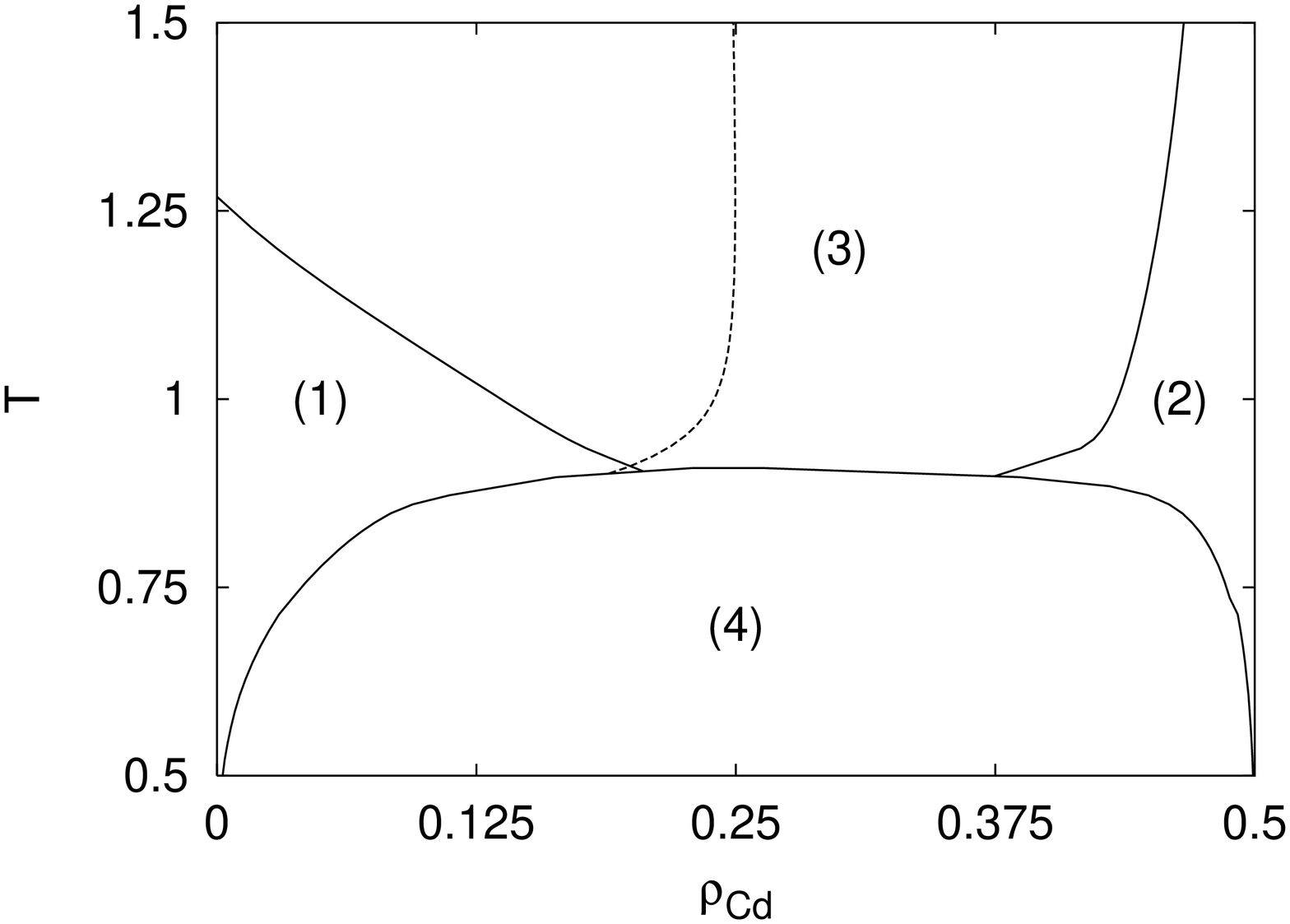}}}}
\put(50,33){\resizebox{0.48\textwidth}{!}{\rotatebox{270}{\includegraphics{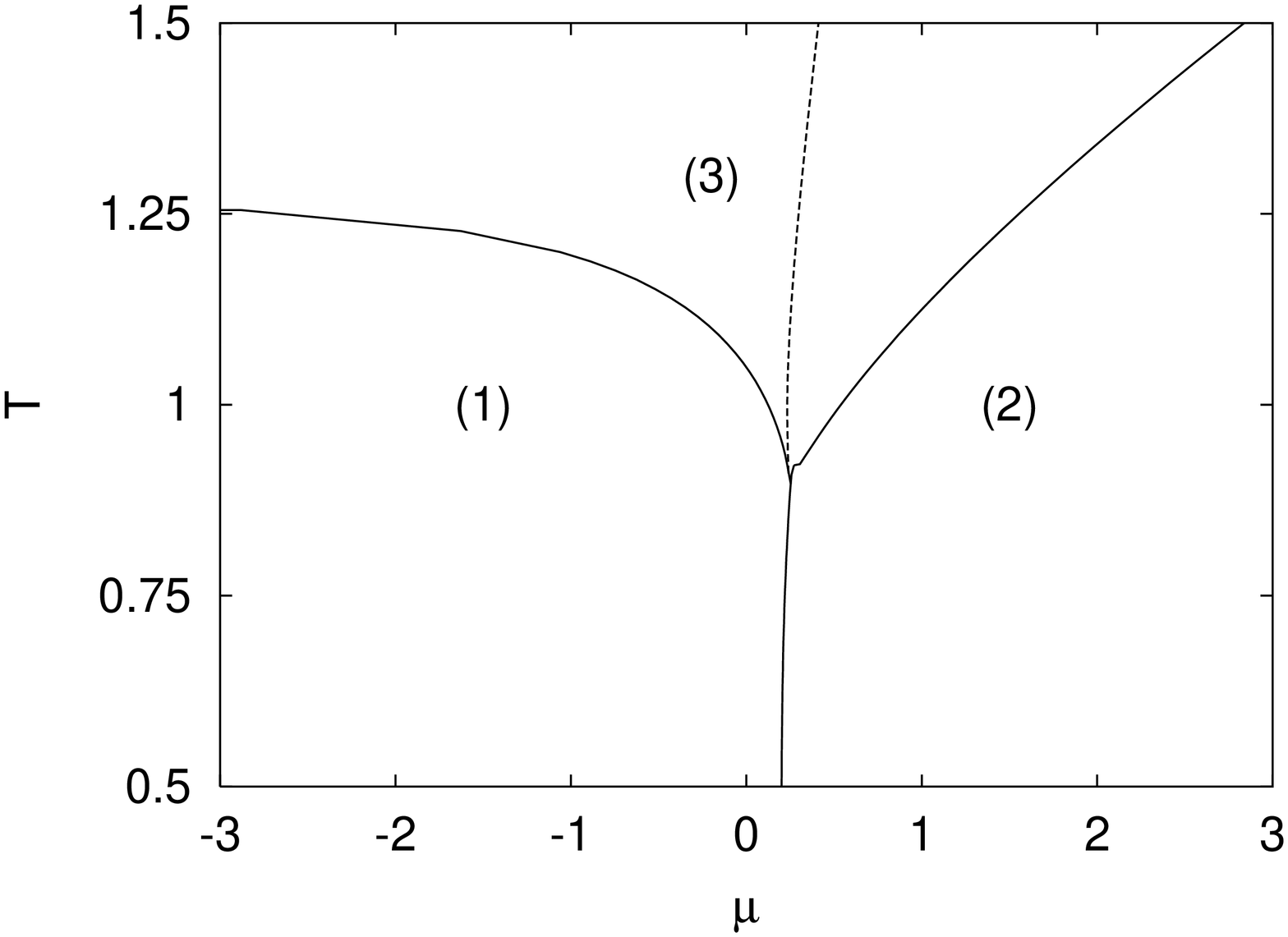}}}}
\put(0,30){(a)}
\put(50,30){(b)}
\end{picture}
\caption{Phase diagram with the parameter set $\epsilon_d = \epsilon_b
= -1$, $\epsilon_x = -1.6$, $\epsilon_t = -1.2$. Compared to the phase
diagram shown in figure \ref{fig2}, the energy difference between the
\cdtxt and \cdtxo reconstruction is greater and the interaction
between dimers is weaker. With these parameters, the transition
between phases (1) and (3) is continuous. Panel (a) shows the phase
diagram in the $\rho_{Cd}$-$T$ plane, in panel (b) it is shown in the
$\mu$-$T$ plane.
\label{fig4}}
\end{figure}

Finally, we discuss the influence of the parameter set on the phase
diagram. The main effect of a variation of the binding energy of the
Te dimers $\epsilon_b$ is to shift the chemical potential at which the
phase transitions occur. The influence on the shape of the phase
diagram is small as long as $\epsilon_b$ is sufficiently large such
that the density of Te atoms which are neither dimerized nor bound to
Cd atoms is small. Due to the electron counting rule, a state where a
large fraction of the Te atoms remains unbound should be irrelevant
for CdTe.

A smaller energy difference $\Delta E$ between a perfect \cdtxt and a
perfect \cdtxo reconstruction increases the tendency of Cd atoms to
arrange in a $2\times1$ order. If $\epsilon_x = -1.95$ and all other
parameters are identical to our standard parameter set, the dashed
line where $C_{Cd}^{x} = C_{Cd}^{d}$ is shifted to a higher $\rho_{Cd}
\approx 0.45$. Additionally, we obtain a higher value of the maximal
Cd coverage in the Te rich phase $\rho_{Cd}^{1,max} = 0.28$, which is
achieved at a lower temperature $T = 0.9$. The temperature $T_t = 0.7$
above which the disordered phase is stable is also slightly lower.  In
general, a greater value of $|\epsilon_x|$ increases both the tendency
of the Cd atoms to arrange in a $(2\times1)$ order in the disordered
phase and the concentration of Cd atoms in the \tetxo reconstructed
phase. The temperature of the order-disorder transition in the Cd rich
phase is lowered. Conversely, a {\em greater} energy difference
between perfect \cdtxt and \cdtxo reconstructed phases alters the
temperature of this transition and leads to a preferential
$c(2\times2)$ arrangement of the Cd atoms in the disordered phase.
 
The main effect of a variation of the interaction $\epsilon_t$ between
Te dimers is to change the properties of the transition between the
phases (1) and (3). As long as $\epsilon_t$ is sufficently high, the
phase diagram is qualitatively similar to that shown in figure
\ref{fig2}. In this case, the main effect of a variation of
$\epsilon_t$ is to shift the temperature $T_c^1 \approx \epsilon_t$
where the \tetxo phase vanishes. However, at low $\epsilon_t$ this
transition becomes a {\em continuous} phase transition without any
coverage discontinuity. As an example, in figure \ref{fig2} a phase
diagram with $\epsilon_t = -1.2$, $\epsilon_b = \epsilon_d = -1$ and
$\epsilon_x = -1.6$ is shown. The coexistence regime (4) vanishes at
$T_t = 0.9$ such that there is no phase separation between the
long-range ordered \tetxo phase and the disordered phase. This
parameter set yields a comparatively high $\Delta E$. Therefore, the
line where $C_{Cd}^{x} = C_{Cd}^{d}$ is at relatively low Cd
coverages.  In contrast to the situation at large $\epsilon_t$, the
chemical potential at which the transition from phase (1) to phase (2)
occurs is not independent of temperature. Instead, the transition is
at slightly greater $\mu$ at higher temperature. In consequence, there
is a small range $0.2 < \mu < 0.27$ where there is a phase transition
from phase (2) to phase (1) if $T$ is increased at constant chemical
potential.

\section{Comparison with experimental results \label{experimentcompare}}
 
The results presented in section \ref{results} suggest an
interpretation of the experimentally observed crossover from a \cdtxt
reconstruction to a \cdtxo reconstruction as an accompanying effect of
an order-disorder phase transition. At low temperature, there is a
long-range ordered Cd-rich phase with a \cdtxt reconstruction. At a
critical temperature, the Cd atoms lose their long-range order and
arrange preferentially in a $(2 \times 1)$ pattern. This picture is
consistent with the experimental observation of small domains in the
\cdtxo reconstruction \cite{ntss00} which indicate a high degree of
disorder.

Strictly speaking, a CdTe surface under vacuum is not in thermal
equilibrium.  At the temperature of the \cdtxt - \cdtxo transition,
sublimation plays an important role. However, in a previous
publication \cite{ab01} we have shown that the basic features of a
simplified version of our model which neglects Te dimerization
\cite{baksv01} are preserved under the conditions of step flow
sublimation. This is the dominant sublimation mechanism for CdTe (001)
\cite{nskts00}. Sublimation is slow enough to permit a {\em local
equilibration} of the terminating layer which justifies the
application of equilibrium thermodynamics.  However, the Cd coverage
$\rho_{Cd}$ is determined by the sublimation process. Therefore,
within the limit of an equilibrium model it is not possible to
calculate the path in the phase diagram that CdTe follows.  

Auger measurements \cite{ntss00} yield $\rho_{Cd} \approx 0.35$ at the
transition temperature in vacuum. This suggests that the system is in
the coexistence regime for a wide range of temperatures. Then, the
behaviour of CdTe in vacuum should be similar to our results at
constant $\rho_{Cd} = 0.35$ (section \ref{canonicalresults}). Electron
diffraction techniques investigate large regions on the surface and
hence yield averages over all coexisting phases. Consequently, it
should be reasonable to compare results of these experiments with
quantities which are averaged over the whole system in our canonical
simulations. As discussed in section \ref{canonicalresults}, there are
two independent effects which lead to a clear dominance of the
arrangement of Cd atoms in rows: the order-disorder transition in the
Cd-rich phase and the fact that an increasing fraction of the Cd atoms
is dissolved in the Te-rich phase. Electrons are diffracted both from
the Cd atoms and the Te dimers which have $(2 \times 1)$ order.  The
superposition of these effects should yield the pronounced $(2 \times
1)$ diffraction peaks which have been observed in
\cite{tdbev94,ntss00}.

In the \cdtxt reconstructed phase, at temperatures well below the
phase transition one frequently finds collective thermal excitations
of adatoms, where a row of several Cd atoms is shifted by one lattice
constant in the y-direction.  Due to the repulsion between Cd atoms in
this direction, this is possible only if one Cd atom per excitation is
missing. An example is shown in figure \ref{fig2}d. This effect has
been observed experimentally by Seehover et. al. \cite{sfjetbd95} at
room temperature using STM microscopy.  There is a striking similarity
between figure 3 in \cite{sfjetbd95} and figure \ref{fig2}d.

The phase diagram of our model explains the properties of the
reconstructions of the CdTe (001) surface under an external particle
flux. If the Cd coverage of the surface is increased by deposition of
Cd, the state of the system moves into regions of the phase diagram
where the Cd atoms arrange preferentially in a $c(2 \times 2)$
pattern. Depending on temperature, this is either the ordered \cdtxt
phase (2) or the region of the disorded phase (3) on the right side of
the dashed line where $C_{Cd}^{d} > C_{Cd}^{x}$. Indeed, experiments
\cite{tdbev94,ct97} have shown that an external Cd flux restores the
\cdtxt reconstruction at high temperatures where a $(2 \times 1)$
order is found under vacuum conditions.  Clearly, a strong Te flux
induces the formation of a long-range ordered \tetxo phase at
temperatures below $T_{c}^{1}$.

On ZnSe, a Zn terminated $(2 \times 1)$ reconstruction has not been
observed yet. Our model offers an explanation of this fact which is
consistent with the results of density functional theory
\cite{gn94,pc94,gffh99}. These calculations have shown, that the
difference in the surface energies per $(1 \times 1)$ surface unit
cell between perfect cation terminated $c(2 \times 2)$ and $(2 \times
1)$ reconstructions in ZnSe is significantly greater than in CdTe
($0.03 \mbox{eV}$ versus $0.008 \mbox{eV}$). In our model, this greater
energy difference corresponds to a smaller value of $|\epsilon_x|$
which shifts the line where $C_{Cd}^{x} = C_{Cd}^{d}$ to smaller 
coverages. Consequently, a \zntxt arrangement dominates in a much
wider range of coverages. 

Wolframm et. al. \cite{wewr00} have measured the locus of the
transition beween the \zntxt reconstruction and the \setxo
reconstruction as a function of temperature and the composition of an
external particle flux by means of reflection high energy electron
diffraction (RHEED). They find that at higher temperature a greater Se
flux is required to obtain a $(2 \times 1)$ diffraction pattern.  At
temperatures above 450 $^{o}$C no \setxo reconstruction could be
observed even under extremely Se-rich conditions. This is reminiscent
of our observation that the anion-rich phase (1) vanishes at a
temperature $T_{c}^{1}$.

Unfortunately, the available experimental data are insufficient for a
systematic fit of the model parameters. However, some rough estimates
show that at least the orders of magnitude are reasonable. As
discussed above, the \cdtxt -\cdtxo transition of CdTe in vacuum
should be at a temperature close to $T_t$. Identifying this with the
experimental value of 570 K, we obtain the value of our energy unit
$|\epsilon_d| \approx 0.06 \mbox{eV}$ in physical units. This yields a
value $\Delta E \approx 0.003 \mbox{eV}$ for the difference in the
surface energies of \cdtxt and \cdtxo, which is about $1/2$ of the
value of $0.008 eV$ which has been obtained by means of DFT
calculations. This shows at least that the qualitative agreement
between experiments and our model has been obtained in a physically
reasonable region of the parameter space.

In our model, phase (1) vanishes at a temperature $T_c^1 \approx
|\epsilon_t|$. Identifying this with the value of $450 ^{o}C$ which
has been measured in experiments on ZnSe, we obtain that $\epsilon_t
\approx 0.06 \mbox{eV}$. This is the same order of magnitude as our
estimate of $|\epsilon_d|$ in CdTe. 

These considerations suggest that it should be possible to obtain
quantitative agreement between experiments and our model both for CdTe
and ZnSe with values of the model parameters on the order of magnitude
of a few ten meV.

\section*{Acknowledgements}

We thank Wolfgang Kinzel and Richard Metzler for stimulating
discussions and a critical reading of the manuscript. M.A. was
supported by the Deutsche Forschungsgemeinschaft.  

\appendix
\section*{Appendix: Algorithms for continuous time Monte Carlo simulations}

We have simulated our model using continuous time Monte Carlo
techniques which greatly improve the computational speed compared to
simpler, Metropolis like algorithms \cite{nb99}. A Markov chain of
states $s$ of the system is constructed with transition rates $r_{s
\rightarrow s'}$ between states $s$ and $s'$. This dynamics converges
to a Gibbs distribution if every possible state of the system can be
reached within a finite number of transition events (ergodicity) and
the rates fulfil a detailed balance condition $r_{s \rightarrow
s'}/r_{s' \rightarrow s} = \exp(-(H(s') - H(s))/T)$. In every time
step, one event is performed which is selected randomly with
probability $p_{s \rightarrow s'} = r_{s \rightarrow s'} / R(s)$,
where $R(s) = \sum_{s'} r_{s \rightarrow s'}$ is the sum of transition
rates of all possible events in state $s$. The physical time interval
$\Delta t (s) = 1/R(s)$ which passes between subsequent events depends
on the state of the system. In the calculation of thermal averages,
the observables measured in state $s$ have to be weighted with $\Delta
t(s)$.  This algorithm requires the knowledge of the rates of all
possible events in the current state of the system. If the
configuration of the system is changed only locally in an event, both
the selection of events and the updates of the rates of events can be
done in $\mathcal{O}(\log \mathcal{N})$ CPU steps using a binary
search tree. Here, $\mathcal{N}$ is the number of possible events.

The application to grand-canonical simulations is straightforward.
There are two possible events per lattice site which change the state
of the site: $x_{i,j} \leftarrow (x_{i,j} + 1) \, \mbox{mod} \, 3$ and
$x_{i,j} \leftarrow (x_{i,j} + 2) \, \mbox{mod} \, 3$. These
simulations can be done with Metropolis-type rates $r_{s \rightarrow
s'} = \mbox{min} \{ 1, \exp[-(H(s') - H(s)/T]\}$ or symmetrical rates
$r_{s \rightarrow s'} = \exp[-(H(s') - H(s))/(2T)]$. Both
possibilities yield identical results.

A canonical simulation requires an algorithm where the number of
particles is fixed. In this case, in every event one particle jumps to
a different site. We choose a nonlocal dynamics where the range of
particle jumps is unlimited. This yields considerably faster
equilibration compared to a Kawasaki dynamics with nearest neighbour
diffusion only. For simplicity, we permit only jumps to a site where
the binding energy of the particle is independent of the state of its
initial site, i.e. we forbid jumps to nearest and next nearest
neighbour sites.  If the particle jumps from site $i$ to site $j$, the
energy difference between the final and the initial state is $\Delta H
= \Delta H_j - \Delta H_i$, where $\Delta H_{x}$ is the energy
difference of the system with site $x$ occupied and empty. The rates
\begin{equation}
r_{i \rightarrow j} = \exp \left[ \left( \Delta H_i - \Delta H_j
\right) / \left( 2 T \right) \right]
\end{equation}
fulfil the detailed balance condition. Then, the probability for a
jump from site $i$ to site $j$ factorizes, i. e.
$$
p_{i \rightarrow j} = p_{i}^- \cdot p_{j}^+ \; \; \; \mbox{where} \;
\; \; p_{x}^\pm = \frac{r_x^\pm}{\sum_x r_x^\pm}.
$$ 
Here, we have introduced the rate for deposition of a particle at site
$x$ ($r_{x}^+$) and for removal of a particle at site $x$
($r_x^-$). $r_x^+ = \exp[-(\Delta H_x)/(2 T)]$ if site $x$ is empty
and zero otherwise. Conversely, $r_x^- = \exp[\Delta H_x/(2 T)]$ on
occupied sites and zero on empty sites. Due to this factorization
property we can proceed in two steps: In the first step, we select
the site $i$ from which the particle starts with probability
$p_{i}^-$. Then, we select the site $j$ where the particle is landing
with probability $p_j^+$. If the distance between site $i$ and site
$j$ is $\geq 2$ the particle is moved. Otherwise, the event is
rejected and the system remains unchanged. Since the number of
rejected events is small on large systems, the loss of speed can be
neglected.

However, in our model only the number of Cd atoms is fixed, while the
number of Te dimers may change. This requires a more elaborate
algorithm which uses both canonical and grand-canonical techniques. A
Cd atom may jump to any site which is not occupied by a Cd atom. If
the arrival site is occupied by a dimer, the dimer is destroyed. The
removal of the Cd atom at the starting site creates a pair of Te
atoms. We consider both the case where these Te atoms dimerize
immediately and the case where they remain unbound.  Additionally, Te
dimers may break up and dimers may be created on empty sites. In the
following, these processes will be denoted as ``dimer flips''. In each
timestep, a random number is drawn to decide whether a Cd jump with
immediate formation of a dimer at the starting site, a Cd jump without
dimerization or a dimer flip will occur. The probabilities are
proportional to the sums of the rates of the processes in each
group. A $L \times N$ system contains $\mathcal{O}(L N)$ Cd atoms and
Te dimers. Thus, there are $\mathcal{O}(L N)$ dimer flip events. Since
there are $\mathcal{O}(L N)$ possible arrival sites for each Cd atom,
there are $\mathcal{O}(L^2 N^2)$ Cd jump events. To keep the ratio of
dimer flips and Cd jumps independent of the system size, the dimer
flips have been weighted with a prefactor $L N$. Then, one event in
the selected group is perfomed. For dimer flips, this is done with the
grand-canonical algorithm while Cd jumps are performed by the
canonical two-step algorithm.

\end{document}